\documentclass[letterpaper,twocolumn,10pt]{article}

\usepackage{usenix,epsfig,endnotes}

\usepackage{url}
\usepackage{amsmath}
\usepackage{amsthm}
\usepackage{graphicx}
\usepackage{subfig}
\usepackage{array}
\usepackage{rotating}
\usepackage{multirow}
\usepackage{colortbl}

\definecolor{kugray5}{RGB}{224,224,224}
\usepackage{listing}
\usepackage{amssymb}
\usepackage{algorithmic}
\usepackage[linesnumbered,noend]{algorithm2e}
\usepackage{placeins}
\usepackage[table]{xcolor}
\usepackage{tikz}
\usepackage{lipsum}
\usepackage{pifont}
\usepackage{comment}
\usepackage[font=small]{caption}
\newcommand{\cmark}{\checkmark}

\newcommand{\change}{\textcolor{black}}
\newcommand{\simd}{\textsc{simd-x}}

\newcommand\crule[3][black]{\textcolor{#1}{\rule{#2}{#3}}}

\SetCommentSty{mycommfont}

\usepackage{titlesec}

\titlespacing\section{0pt}{10pt plus 0pt minus 2pt}{4pt plus 0pt minus 2pt}
\titlespacing\subsection{0pt}{4pt plus 0pt minus 2pt}{2pt plus 0pt minus 2pt}

\title{{\simd}: Programming and Processing of Graph Algorithms on GPUs}
\author{
{\rm Hang Liu}\\
University of Massachusetts Lowell 
\and 
{\rm H. Howie Huang}\\
The George Washington University
} 

\begin{document}

%
%
\newcommand\encircle[1]{%
	\tikz[baseline=(X.base)] 
	\node (X) [draw, shape=circle, inner sep=0, fill=black, text=white] {\strut #1};}
%

\maketitle


\section*{Abstract}
\noindent
With high computation power and memory bandwidth, graphics processing units (GPUs) lend themselves to accelerate data-intensive analytics, especially when such applications fit the {single instruction multiple data} (SIMD) model.
However,  graph algorithms such as breadth-first search and k-core, often fail to take full advantage of GPUs, due to irregularity in memory access and control flow. 
To address this challenge, we have developed {\simd}, for programming and processing of {single instruction multiple, {complex}, data} on GPUs.
Specifically, the new {Active-Compute-Combine} (ACC) model not only provides ease of programming to programmers, but more importantly creates  opportunities for system-level optimizations. 
{
To this end, {\simd} utilizes {just-in-time task management} which  filters out inactive vertices at runtime and intelligently maps various tasks to different amount of GPU cores in pursuit of workload balancing.}
In addition, {\simd} leverages {push-pull based kernel fusion} that, with the help of a new deadlock-free global barrier, reduces a large number of computation kernels to very few. 
Using {\simd}, a user can program a graph algorithm in tens of lines of code, while achieving {3$\times$, 6$\times$, 24$\times$, 3$\times$ speedup over Gunrock, Galois, CuSha, and Ligra, respectively.}
\section{Introduction}
\label{sec-introduction}


The advent of big data exacerbates the need of extracting useful knowledge within an acceptable time envelope.
For performance acceleration, many applications utilize graphics processing units (GPUs) whose huge success comes from exploiting the data-level parallelism in these applications. 
{
Implicitly, the traditional single instruction multiple data (SIMD)  model of GPUs assumes {regular} programming and processing, that is, not only the same instruction is executed but also the same amount of work is expected to perform on each piece of data.
Unfortunately, neither assumption holds true for many emerging {irregular applications}, especially graph analytics which is the focus of this work.
That is, such applications do not conform to the  SIMD model, where different amount of work, or worse, completely different work, need to be performed on the data in parallel.
}

To enable graph computation on GPUs, this work advocates a new parallel framework, {\simd}, for the programming and processing of \textit{single instruction multiple, {complex}, data} on GPUs. 
At the heart of {\simd} is the decoupling of programming and processing, that is, {\simd} utilizes the \textit{data-parallel} model for ease of expressing of graph applications, while enabling system-level optimizations at run time to deal with the \textit{task-parallel} complexity on GPUs.
{With {\simd}, a programmer simply needs to define what to do on which data, without worrying about the issues arisen from irregular memory access and control flow, both of which prevent GPUs from achieve massive parallelism. }




\begin{figure*}[!thp]
	\centering
	\includegraphics[scale=0.4]{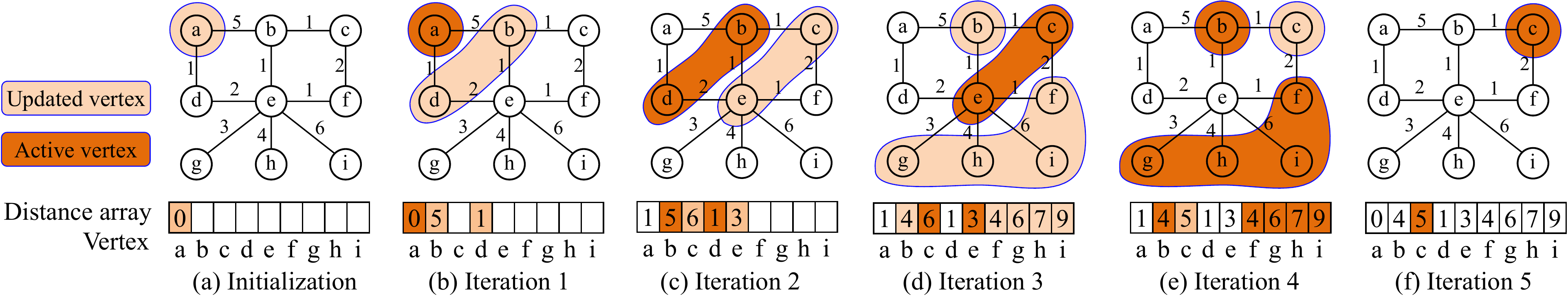}
	\caption{SSSP on a graph, with nine vertices \{a, b, c, d, e, f, g, h, i\} and ten undirected edges (with weights). SSSP iteratively computes on the graph and generates the distance array. Particularly,  heavy and light shadows represent active and most recently updated vertices, respectively.} 
	\label{back-sample-sssp}
		\vspace{-0.1in}
\end{figure*} 


{\simd} consists of three major components: 
First, {\simd} utilizes a new {Active-Compute-Combine} (ACC) programming model that asks a program to define three data-parallel functions: the condition for determining an \textit{active} vertex, \textit{computation} to be performed on an associated edge, and \textit{combining} the updates from edge compute to vertex state.
As we will show later, ACC is able to support a large variety of graph algorithms from breadth-first search, k-core, to belief propagation.
{While ACC adopts the Bulk Synchronous Parallel (BSP) model, it differs from traditional CPU-based graph abstractions such as edge- or vertex-centric models in that ACC avoids atomic operation, enables collaborative early termination (for BFS) and fine-grained task management on GPUs.}
  

Second,  {\simd} relies on just-in-time (JIT) task management to balance parallel workloads across different GPU cores with minimal overhead.
A good task list can increase not only parallelism, but also sequential memory access for the computation of next iteration, both of which are crucial for high-performance computing on GPUs. 
To this end, we have designed a set of new task management mechanisms of online and ballot filters, each of which excels at the complementary scenarios, i.e., the former favors a small amount of tasks while the latter larger tasks.
At runtime, {\simd} judiciously selects the more suitable filter to assemble the active work list for the next iteration. 
{Our JIT task management can largely reduce the memory consumption, thereby accommodate the  graphs much larger than prior work~\cite{merrill2012scalable,wang2015gunrock}.}
Moreover, {\simd} is able to deliver 16$\times$ speedup, on average, across various algorithms and graphs. 

Third, {\simd} designs a new technique of push-pull based kernel fusion which aims to further accelerate graph computing by reducing kernel invocation overhead and global memory traffic. 
{{\simd} addresses the deadlock issue which occurs, when fusing kernel across the software global barrier~\cite{xiao2010inter}, in existing work such as Gunrock~\cite{wang2015graphq}, CuSha~\cite{khorasani2014cusha}, and StreamScan~\cite{yan2013streamscan}. }
Besides, instead of aggressively fusing the algorithm into one giant kernel, {\simd} fuses the  kernels around the pull and push stages within each computation {to minimize both register consumption and kernel relaunching}.
The evaluation shows that the new fusion technique can reduce the register consumption by half and thus double the configurable thread count, leading to 42\% and 25\% performance improvement over non-fused and aggressive fusion, respectively. 


{\simd} is different from prior work in several aspects. 
First, in order to use existing systems efficiently, a programmer needs to possess an in-depth knowledge of GPU architecture~\cite{gaster2012can, nvidia2011nvidia}, e.g., Gunrock requires explicit management of GPU threads and memory~\cite{wang2015gunrock}, and B40C~\cite{merrill2012scalable} and Enterprise~\cite{hang2015enterprise} need thousands of lines of CUDA code for BFS specific optimizations. 
One of the goals of this work is to provide a simple programming model and delegate the responsibility of task management to {\simd}. 
Second, current systems either ignore workload imbalance as in ~\cite{khorasani2014cusha, zhong2014medusa}, or resolve it reactively as in ~\cite{wang2015gunrock, tzeng2010task}, both of which result in undesired system performance.
Lastly, because GPUs lack support for global synchronization, existing systems~\cite{wahib2014scalable,wang2015gunrock,hang2015enterprise,liu2016ibfs,sorensen2016portable} either rely on the multi-kernel design or runtime tunning, both of which come with considerable overhead, especially for graph algorithms with high iteration count. 
{\simd} addresses these challenges with the help of new filters, as well as a deadlock-free software  barrier.


The rest of this paper is organized as follows: 
Section~\ref{sec-back-obj-challenge} presents the challenges of constructing {\simd} on GPUs.
Section~\ref{sec-model} describes the ACC model.
Section~\ref{sec-sched} presents our just-in-time management approach and Section~\ref{sec-fuse} discuses the kernel fusion design. 
We present the graph algorithms in Section~\ref{sec-app} and the evaluation results in Section~\ref{sec-exp}. 
Section~\ref{sec-related} discusses the related work and Section~\ref{sec-conclusion} concludes.

\section{{\simd} Challenges and Architecture}
\label{sec-back-obj-challenge}

\subsection{Graph Computing on GPUs}
\label{sec-graph-compute}

\begin{figure}[t]
	\centering
	\includegraphics[scale=0.4]{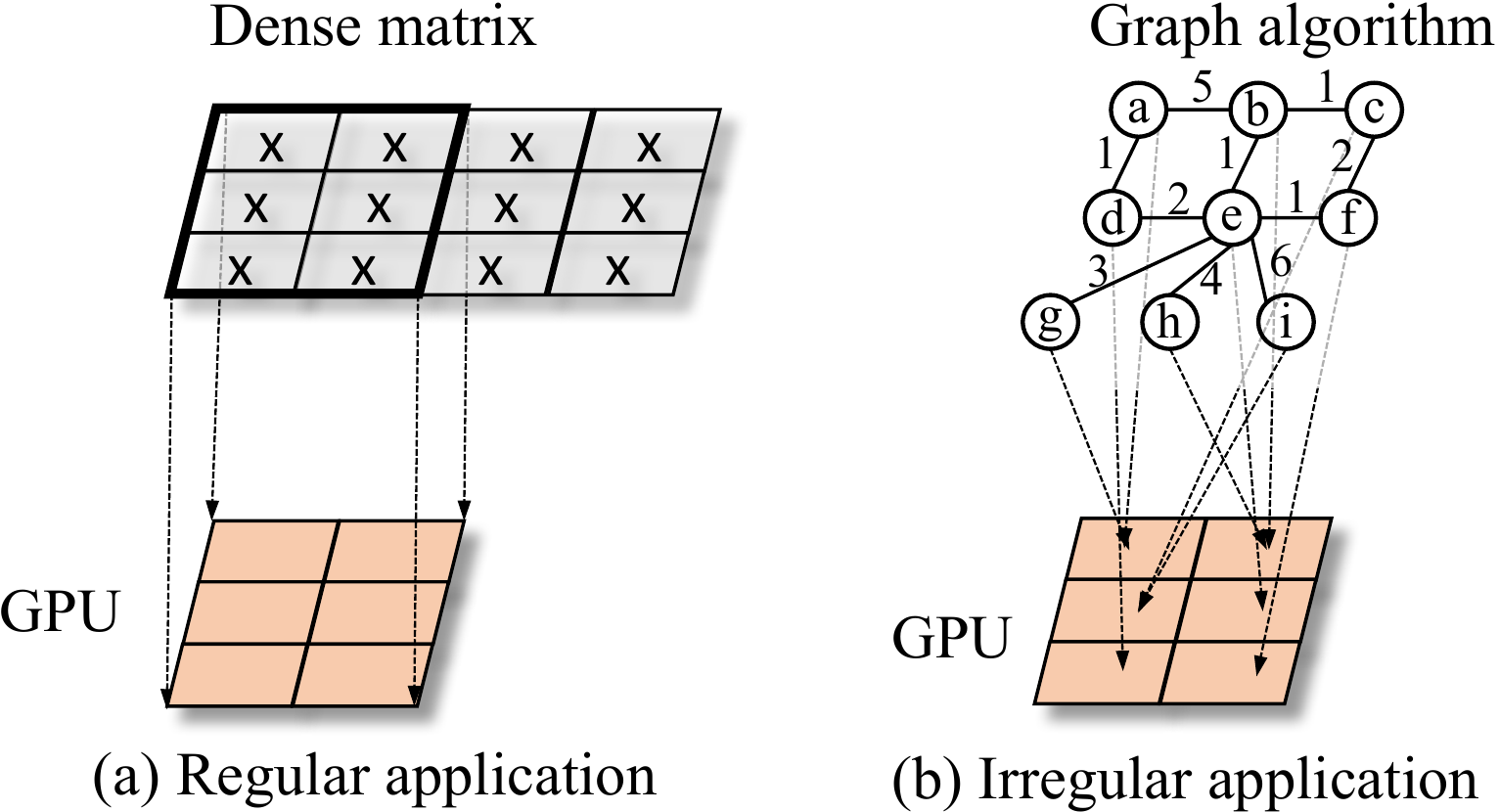}
	\caption{Mapping regular versus irregular applications on GPUs}
	\label{back-reg-irreg}
		\vspace{-0.2in}
\end{figure}

Generally speaking, regular applications present uniform workload distribution across the data set.
As a result, such applications lend themselves to the data-parallel GPU architecture.
For development and evaluation, this work mainly uses NVIDIA GPUs, which have tens of streaming processors and in total thousands of {Compute Unified Device Architecture} (CUDA) cores~\cite{nvidia2011nvidia,nvidia2013kepler}. 
Typically, a \textit{warp} of 32  threads execute the same instruction in parallel on consecutive data. 
For regular application, programming and processing is simple, e.g., dense matrix algebra as shown in Figure~\ref{back-reg-irreg}(a).

On the other hand, task management for irregular applications is challenging on GPUs. 
In this work, we focus on a number of graph algorithms such as breadth-first search, k-core, and belief propagation. 
Here we use one algorithm -- Single Source Shortest Path (SSSP) -- to illustrates the challenges. 
Simply put, a graph algorithm computes on a graph $G$ = ($V$, $E$, $w$), where $V$, $E$ and $w$ are the sets of vertices, edges, and edge weights. 
The computation updates the algorithmic metadata which are the states of vertices or edges in an iterative manner. 
A typical workflow of SSSP is shown in Figure~\ref{back-sample-sssp}. 
Initially, SSSP assigns the infinite distance to each vertex in the \textit{distance array}, which is represented as blank in the figure.
Assuming the source vertex is $a$, the algorithm assigns 0 as its initial distance, and now vertex $a$ becomes active.
Next, SSSP computes on this vertex, that is, calculating the updates for all the neighbors of vertex $a$. 
In this case, vertices \{$b$, $d$\} have their distances updated to 5 and 1 in the distance array. 
At the next iteration, the vertices with newly updated distances become active and perform the same computation again. 
This process continues until no vertex gets updated. 
Different from breadth-first search, SSSP may update the distances of some vertices across multiple iterations, e.g., vertex $b$ is updated in iteration 1 and 3. 

In this example, not every vertex is active at all time, and vertices with different degrees (number of edges) yield varying amounts of workloads. For instance, at iteration 3 of Figure~\ref{back-sample-sssp}(d), one thread working on vertex $c$ computes two neighbors, while another thread on vertex $e$ four neighbors.
As a result, a complex mapping as shown in Figure~\ref{back-reg-irreg}(b) is required for high-performance processing, and to do so necessitates in-depth knowledge from a programmer on GPUs. 

\subsection{Architecture}
\label{sec-challenge}



\begin{figure}[t]
	\centering
	\includegraphics[scale=0.4]{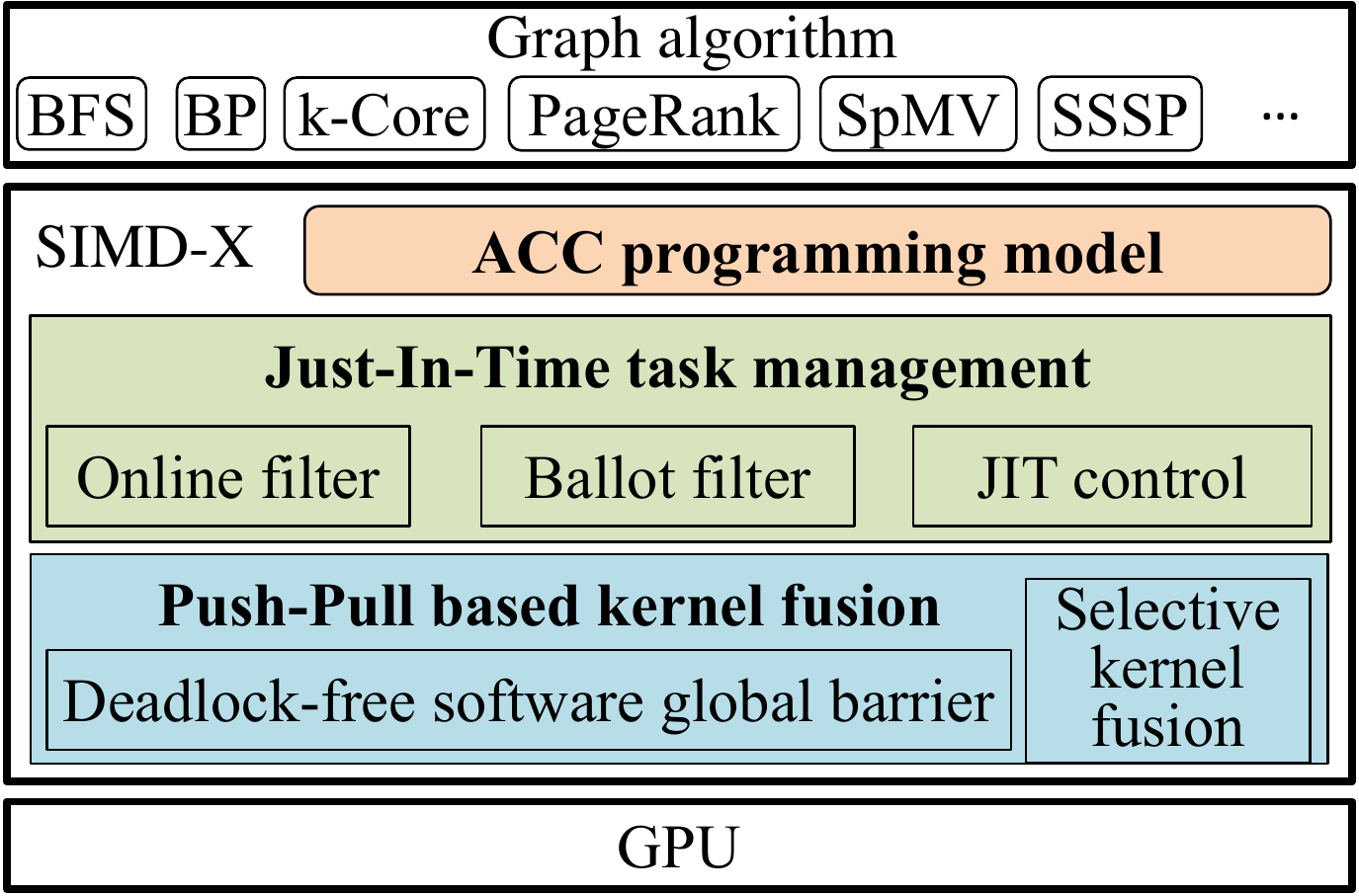}
	\caption{{\simd} architecture}
	\label{frame-arch}
	\vspace{-0.1in}
\end{figure}

{\simd} is motivated to achieve two goals simultaneously: providing ease of programming for a large variety of graph algorithms, whereas enabling fine-grained optimization of GPU resources at the runtime.
Figure~\ref{frame-arch} presents an overview of {\simd} architecture.
To achieve the first goal, {\simd} utilizes a simple yet powerful {Active-Compute-Combine} (ACC)  model.
This data-parallel API allows a programmer to implement graph algorithms with tens of lines of code (LOC).
Prior work requires significant programming effort~\cite{merrill2012scalable, hang2015enterprise, wang2015gunrock}, or runs the risk of  poor performance~\cite{khorasani2014cusha}.

In {\simd}, high-performance graph processing on GPUs is achieved through the development of two components:
(1) JIT task management, which is responsible for translating data-parallel code to parallel tasks on GPUs.
Essentially, {\simd} ``filters" the inactive tasks and groups similar ones to run on the underlying SIMD architecture.
In particular, {\simd} develops online and ballot filters for handling different types of tasks, and dynamically selects the better filter during the execution of the algorithm.
And (2) Pull-push based kernel fusion. 
Graph  applications are iterative in nature and thus require synchronizations. Fusing kernels across iterations would yield indispensable benefits, because kernel launching at each iteration incurs non-trivial overhead. 
In {\simd}, we observe that with aggressive kernel fusion,  register consumption would increase  dramatically, lowering the occupancy and thus performance. 
To this end, {\simd} deploys kernel fusion around pull and push stages of each graph computation, seeking a sweet spot that not only maximizes the range of each kernel fusion  but also minimizes the register consumption. 
It is  worthy noting that we also address the deadlock issue faced by  software global barrier  in {\simd}.

\section{ACC Programming Model}
\label{sec-model}

When it comes to graph computing, there are two main programming models: vertex-centric vs. edge-centric.
``Think like a vertex"~\cite{malewicz2010pregel,zheng2015flashgraph} focuses on tasks on active vertices in a graph, whereas ``think like an edge"~\cite{roy2013x,roy2015chaos} iterates on edges and simplifies programming.
{\simd} aims to achieve the dual goal of ease of programming in edge-centric model, and efficient workload scheduling in vertex-centric model.


\begin{table*}[]
	\centering
	\caption{\small Comparison between ACC and relevant GPU-based programming models. \crule[green]{.18in}{.1in} denotes desirable feature.}
	{\scriptsize
		\label{tab-related}
		\begin{tabular}{|l|l|l|l|l|l|}
			\hline
			&  & \multicolumn{3}{c|}{Stages} &  \\ \cline{3-5}
			\multirow{-2}{*}{Abstraction} & \multirow{-2}{*}{Related Work} & Task filtering & Workload balancing & Avoid atomic operation & \multirow{-2}{*}{Graph format} \\ \hline
			ICU & CuSha~\cite{khorasani2014cusha}, Lux~\cite{jia2017distributed} &  & \cellcolor[HTML]{67FD9A}Init/Compute (Edge) & \cellcolor[HTML]{67FD9A}Update & Edge list \\ \hline
			ICRU & WS~\cite{khorasani2015scalable}  &  & \cellcolor[HTML]{67FD9A}Init/Compute (Edge) & \cellcolor[HTML]{67FD9A}Reduce/IsUpdate & \cellcolor[HTML]{67FD9A}CSR \\ \hline
			AFC & Gunrock~\cite{wang2015gunrock} & \cellcolor[HTML]{67FD9A}Advance/Filter & Compute (Vertex, with atomic update) &  & \cellcolor[HTML]{67FD9A}CSR \\ \hline
			GAS & GTS~\cite{kim2016gts}, GraphReduce~\cite{sengupta2015graphreduce} &  & \cellcolor[HTML]{67FD9A}Gather (Edge) & Apply/Scatter & Edge list \\ \hline\hline
			ACC & SIMD-X & \cellcolor[HTML]{67FD9A}Active & \cellcolor[HTML]{67FD9A}Compute (Edge) & \cellcolor[HTML]{67FD9A}Combine & \cellcolor[HTML]{67FD9A}CSR \\ \hline
		\end{tabular}
	}
\vspace{-.2in}
\end{table*}

\subsection{Motivation}
\label{subsec-motivation}

{
\noindent\textbf{Graph programming} converges to either \textit{vertex-centric} or \textit{edge-centric} models. In particular, the {vertex-centric model} contains two functions:  
\textsf{vertex\_scatter} defines what operations should be done on this vertex, and \textsf{vertex\_gather} applies the updates on the vertex. This model has been adopted by a number of existing projects, e.g., Pregel~\cite{malewicz2010pregel}, GraphLab~\cite{low2010graphlab}, PowerGraph~\cite{gonzalez2012powergraph}, GraphChi~\cite{kyrola2012graphchi}, FlashGraph~\cite{zheng2015flashgraph}, Mosaic~\cite{maass2017mosaic}, and GridGraph~\cite{zhu2015gridgraph}, as well as GPU-based implementation such as CuSha~\cite{khorasani2014cusha} and Gunrock~\cite{wang2015gunrock}. 
On the other hand, the {edge-centric model} is initially introduced by the external-memory graph engine X-stream~\cite{roy2013x} to improve IO performance.
It requires a programmer to define two functions needed on each edge, \textsf{edge\_scatter} and \textsf{edge\_gather}.  
As such, this model schedules threads by the edge count. Particularly, one thread  needs to send the information of the source vertex and the outbound edge to the destination vertex (\textsf{edge\_scatter}), which atomically applies the new updates in \textsf{edge\_gather}.
}

{
In this work, we believe the many-threaded nature of GPU architecture demands a new abstraction. We intend to exploit various thread scheduling options to better tackle workload imbalance~\cite{hang2015enterprise,wang2015gunrock}, while minimizing the overhead with regards to  atomic operations on GPUs~\cite{luo2010effective}. 
Table~\ref{tab-related} summarizes the designs of recent GPU-based graph analytics systems. To avoid wasting the threads to compute on inactive vertices, \textit{task filtering} is essential in generating a list of active vertices. Once task lists are ready, \textit{workload imbalance} caused by skewed degree distribution in many graphs becomes the next concern. Since handling this issue in a vertex centric model involves nontrivial programming efforts~\cite{hang2015enterprise},  edge-based computing presents a desirable alternative. However, traditional edge-centric approach would result in atomic updates at the destination vertex, thus a proper schedule before applying the update is essential to \textit{avoid atomic operation}. It is also important to note that \textit{compressed sparse row} (CSR) is a preferable graph format which can save around 50\% of the space over edge list format, as contemporary GPUs only feature tens of GB memory~\cite{nvidia2011nvidia}. The proposed ACC framework is  designed to address these three challenges.
}

\subsection{ACC Model}

The new ACC model contains three functions: \textbf{A}ctive, \textbf{C}ompute, and \textbf{C}ombine. 
ACC supports a wide range of graph algorithms and  requires much fewer lines of code compared to prior work. 
In this following, we will discuss the three functions.

\begin{figure}[!ht]
	\centering
	\includegraphics[scale=0.6]{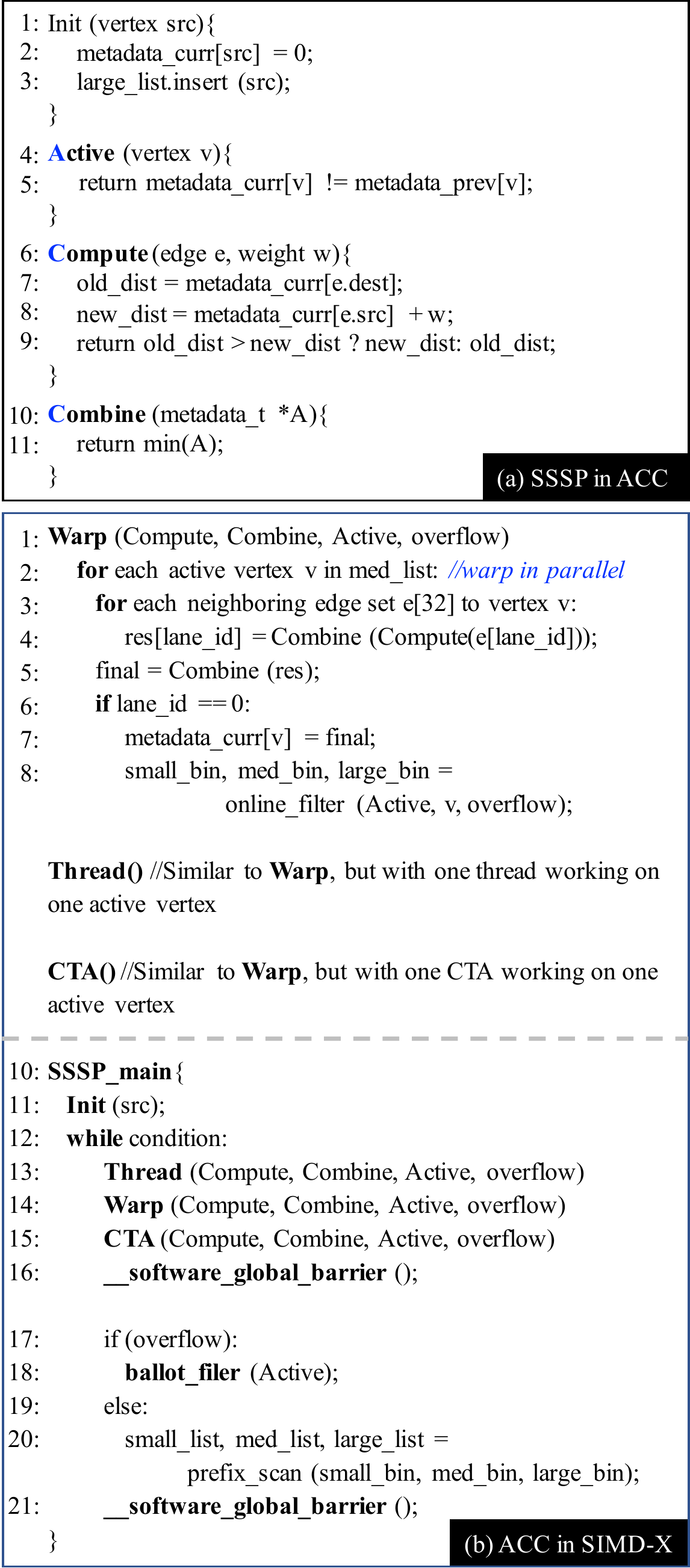}
	\caption{{(a) Single-Source Shortest Path expressed in ACC model and (b) ACC abstraction within {\simd} framework.}} 
	\label{tech-compute}
	\vspace{-.2in}
\end{figure}

\vspace{2mm}
\noindent \textbf{Active} allows a programmer to specify the condition whether a vertex is active.
Formally it can be defined: $$\exists_v \leftarrow active (M_v, v)$$ where $v$ is the vertex ID and $M_v$ represents its metadata. 
Depending on the algorithm, the \textsf{Active} function may vary. Belief propagation (BP) is simple which treats all vertices as active. In comparison, SSSP, as shown in Figure~\ref{tech-compute}(a), considers the vertices active when their current metadata differs from the prior iteration. 


Simply put, {\simd} distinguishes active vertices from inactive ones, and focuses on the calculation needed for each vertex. 
This is different from the vertex-centric model which deals with not only the active vertex but also its neighbors. 
Because two vertices may have different numbers of neighbors, existing systems~\cite{malewicz2010pregel,gonzalez2012powergraph}  likely suffer from workload imbalance.
To this end, {\simd} leverages a classification technique, similar to Enterprise~\cite{hang2015enterprise}, to group the active vertices depending on the expected workload. 

\vspace{2mm}
\noindent \textbf{Compute} defines the computation that happens on each edge. In particular, it specifies the operations on the metadata of edge $(v,\ u)$ and two vertices $v$ and $u$, which can be written as follows: 
$$update_{v\rightarrow u} \leftarrow compute (M_v, M_{(v, u)}, M_u)$$
where the return value of  $update_{v\rightarrow u}$ will be used by the  \textsf{Combine} function. 
For example, SSSP computes the updated distance for the destination vertex as shown in Figure~\ref{tech-compute}(a).


\vspace{2mm}
\noindent \textbf{Combine} merges all the updates, once  the computations are completed. It can be represented:
$$update_u \leftarrow \underset{{v\in Nbr[u]}}{\oplus} update_{v\rightarrow u}$$
where  $\oplus$   must be commutative and associative, e.g., sum and minimum, and is being applied to all the neighbors of vertex $u$. Figure~\ref{tech-compute}(a) presents the \textsf{Combine} examples of SSSP. Particularly, BP summarizes all updates, where SSSP combines all updates from compute by selecting the minimum. 

{\simd} optimizes two types of combine operations, i.e., aggregation and voting. Particularly, aggregation cannot tolerate overwrites, that is, all updates are needed for computing the results. PageRank, SSSP and k-Core are representative examples of such operation. 
In contrast, voting relaxes this condition, that is, the algorithm is correct as long as one update is received because all updates are identical. 
BFS, weakly connected component and strongly connected component algorithms~\cite{slota2014bfs} fall into this category.
%

\subsection{Processing with ACC}
\label{sec-api}

{
This section uses SSSP an example to illustrate how the {\simd} framework works. 
SSSP computes the shortest paths between the source vertex and the remaining vertices of the graph. Although similar to Breadth-First Search (BFS), SSSP is more challenging as only one vertex with the shortest distance should be computed at one time. 
To improve the parallelism, we adopt the delta-step~\cite{meyer1998delta} algorithm which permits us to simultaneously compute a collection of the vertices whose distances are relatively shorter. We assume positive edge weights.}

{
As shown in line 12 - 21 of Figure~\ref{tech-compute}(b), {\simd} structures graph computation as a loop. Similar to popular GPU-based frameworks~\cite{wang2015gunrock,khorasani2014cusha,khorasani2015scalable}, ACC follows BSP model, that is, synchronization is required at the end of each iteration. As we will discuss in the next section, {\simd} employ three kernels to balance the workload, Thread, Warp and CTA kernels working on small\_list, med\_list and large\_list, respectively. During computing, the online\_filter (Section~\ref{sec-sched}) attempts to track the active vertices with the thread bins (i.e., small\_bin, med\_bin and large\_bin). Note that each active vertex is stored in one of these three bins based upon its degree. After a deadlock free software global barrier (Section~\ref{sec-fuse}), {\simd} checks whether an overflow happens in any of the thread bins, which leads to either a ballot filter-based active lists generation or a simple prefix-scan based concatenation of all thread bins to produce the active lists (line 17-21).
}

\begin{figure}[t]
	\centering
	\includegraphics[scale=0.45]{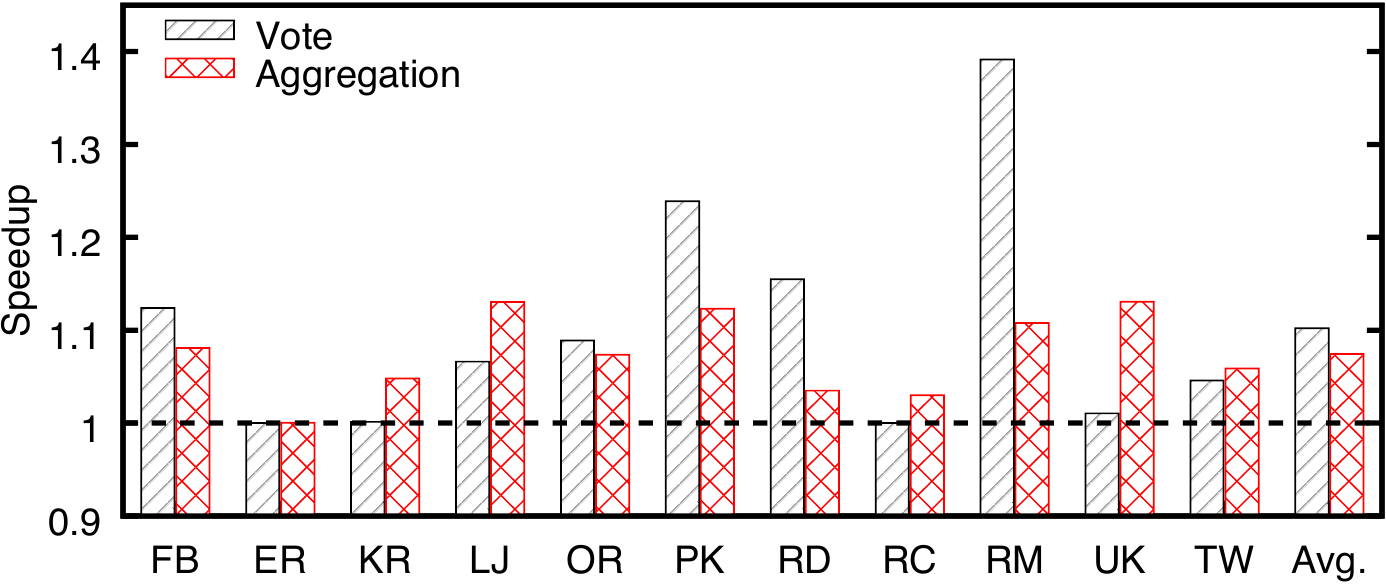}
	\caption{Speedup of our ACC model over Gunrock. Note vote and aggregation operations are materialized by BFS and SSSP algorithms, respectively.} 
	\label{acc-vs-compute-atomic}
	\vspace{-0.2in}
\end{figure}

{
In Figure~\ref{tech-compute}(b), Line 1 - 8 exemplifies the interactions between ACC and {\simd}. Firstly, {\simd} will schedule a warp of threads to work on the neighbors of one active vertex from {med\_list}. Similarly, Thread and CTA will schedule a thread and CTA to work on each active vertex from small\_list and large\_list, respectively. During computation, each thread will conduct a local Compute and Combine at line 4. Once finished, a cross Warp Combine happens at line 5. Eventually, the first thread from the Warp applies the final updates (without atomic operation) and store this vertex (if active) into corresponding thread bins.
}

\begin{figure*}[t]
	\centering
	\includegraphics[scale=0.235]{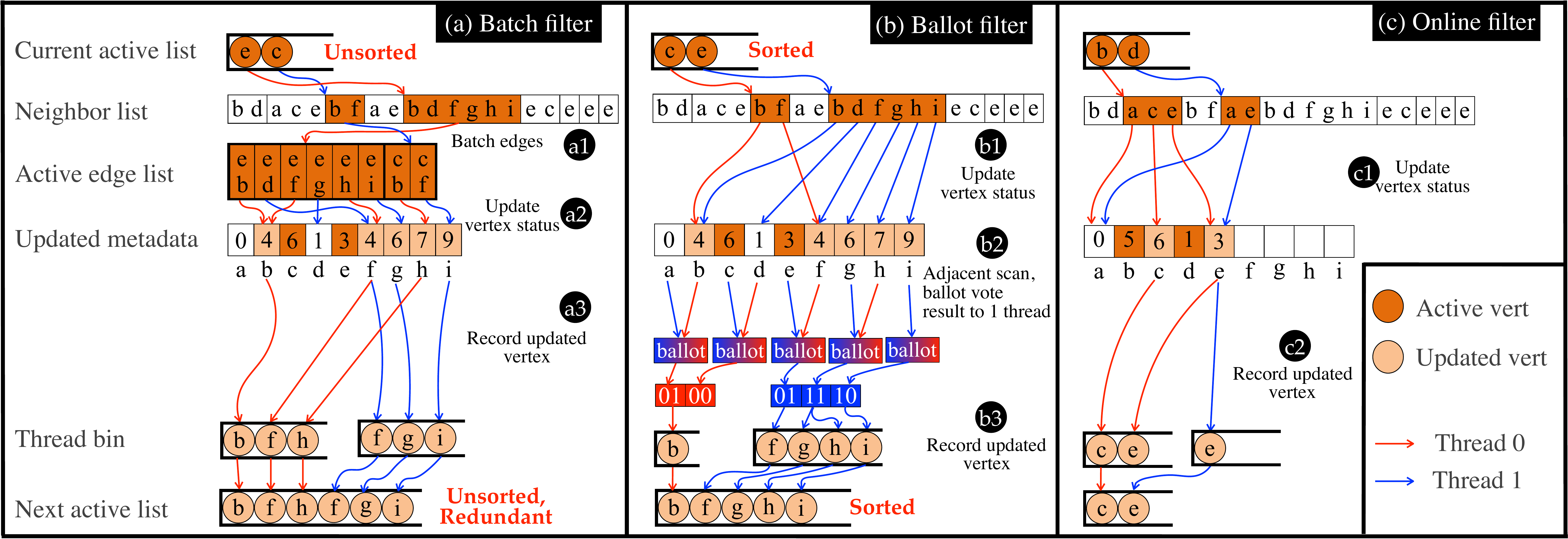}
	\caption{Three task management methods. Particularly, batch filter and ballot filter work on Figure~\ref{back-sample-sssp}(d) to produce a task list for next iteration. Online filter does that for Figure~\ref{back-sample-sssp}(c).} 
	\label{tech-filter-example}
	\vspace{-.2in}
\end{figure*}

\vspace{2mm}
\noindent
\textbf{Comparison}
The new ACC model follows a computation and then combine approach which pays the extra  overhead (i.e., assembling all updates residing in shared memory from participating threads) in order to achieve the benefits of atomic-free updates. 
Gunrock, in contrast, directly applies computation update to vertex status with atomic operations, thereby avoids inter-thread communication but experiences heavier overhead from atomic operation.
Figure~\ref{acc-vs-compute-atomic} studies the performance impact of ACC vs Gunrock.  
One can see that ACC is, on average, 12\% and 9\% faster on vote and aggregation operations, respectively. For vote, the speedup comes from the fact that ACC can schedule all threads to collaboratively determine early termination, which is not possible in Gunrock. For aggregation,  the performance gains comes from the elimination of atomic updates.

\section{Just-In-Time Task Management}
\label{sec-sched}

Task management is essential for graph applications.
The key to success is to ensure good workload balance on GPUs,  that is, each GPU core, regardless of from which streaming processor, accounts for a similar amount of workload.
{To this end, we adopt a two-step approach -- task management and thread assignment. In \textbf{step I: task management}, the tasks are classified into various lists, namely {small\_list, med\_list and large\_list}. In \textbf{step II: thread assignment}, various granularity of GPU threads are scheduled to work different worklists. That is, a single thread per small task, a warp per medium task and a CTA per large task.
Unlike prior work~\cite{hang2015enterprise,wang2015gunrock,merrill2012scalable}, {\simd} focuses on improving the first step as it is the major culprit that offsets the benefits of workload balancing.} 
In the following, we will first analyze the drawback of existing batch filter method, then describe our proposed two new filters, as well as the JIT selection mechanism.

\vspace{2mm}
\noindent \textbf{Drawback of batch filter.} This approach~\cite{wang2015gunrock,merrill2012scalable,davidson2014work} first loads all the edges of the active vertices to construct an active edge list. 
Still using the example of SSSP in Figure~\ref{back-sample-sssp}(c), this step loads neighbors of vertex \{e, c\} and constructs the active edge list in {\scriptsize \encircle{a1}} of Figure~\ref{tech-filter-example} (a). 
Next, batch filter checks these edges and updates vertex metadata {\scriptsize \encircle{a2}}, followed by recording the updated vertices in thread bin at step {\scriptsize \encircle{a3}}. Eventually, batch filter will concatenate these thread bins to arrive at a potentially \textit{unsorted and redundant} next active list -- \{$b$, $f$, $h$, $f$, $g$, $i$\}. 
Note, thread private local storage -- thread bin -- is used to avoid the expensive atomic operations, because multiple threads would need atomic operation to put active vertices directly into next active list.

We observe several drawbacks when using the batch filter for various graph algorithms. 
First, the active list can consume up to 2$\cdot|$E$|$ memory space because majority of the vertices in a graph can become active at one iteration~\cite{beamer2012direction, hang2015enterprise}, which is especially true for popular social and web graphs. 
Considering GPU has very limited on-board memory (e.g., 16 GB), this restriction makes large-scale GPU-based graph computing intractable. 
Second, batch filter produces a worklist with unsorted and redundant active vertices, e.g., next active list -- \{$b$, $f$, $h$, $f$, $g$, $i$\} of Figure~\ref{tech-filter-example}(a), which will lead to poor memory performance for next iteration computation. 

\vspace{2mm}
\noindent \textbf{Ballot filter} is designed to overcome all these shortcomings.  
It first loads the neighbors of active vertices and immediately updates vertex metadata. 
As shown at step {\scriptsize \encircle{b1}} in Figure~\ref{tech-filter-example}(b), the neighbors of \{e, c\} get updated immediately. 
Afterwards, thread 0 and 1 (red and blue lines) will exploit ballot scan to inspect the updated metadata and record those updated vertices in local thread bin at step {\scriptsize \encircle{b3}}. The eventual step is similar to batch filter -- we will concatenate these two thread bins to arrive at the next active list, whereas, with \textit{sorted and unique} active vertices. 

Ballot scan is the key to comprehend why we arrive at a better next active list. In steps {\scriptsize \encircle{b2}} and {\scriptsize \encircle{b3}} of  Figure~\ref{tech-filter-example}(b), threads 0 and 1 perform coalesced scan of vertex metadata, and with 
the CUDA $\_\_ballot()$ primitive, return a bit variable `$01$' to the first thread. 
Here $1$ means active and $0$ otherwise, in this case, vertex $a$ is not active while $b$ is.
Through collaboratively working on the entire metadata array, the first thread eventually gets the bit value `0100' for the first four vertices, while the second thread `011110' for the remaining six vertices.  
Consequently, this approach produces a sorted active list, that is, \{$b$, $f$, $g$, $h$, $i$\} in {\scriptsize \encircle{b3}}. 


We intentionally schedule thread 0 and 1 to collaboratively scan the metadata in order to achieve coalesced memory access during scan, as well as, making thread 0 and 1 account for a continuous range of vertices, that is, vertices $a$ - $d$ to thread 0 and $e$ - $i$ to thread 1.  This achieves the dual benefits: coalesced scan and sorted active vertices in next active list. Last but not the least, this scheduling lends ballot filter to be many-thread safe. 

{We also notice an unpublished parallel efforts from Khorasani's dissertation~\cite{khorasani2016high} which is closely related to ballot filter. However, his design relies on atomic operation to compute the offsets of active vertices from each Warp in the next active list and subsequently assigns merely a single thread from the Warp to enqueue all these active vertices. This design implies twin disadvantages comparing to ours. First, atomic operation-based offset computation cannot yield sorted active lists. Second, single thread-based active vertices recording tends to be slower than Warp-based one which is our design.}

Ballot filter is not without its own issue, especially when the amount of active vertices is low. In that case, scanning the metadata array would account for the majority of the runtime. For instance, in ER and RC graphs, 99.23\% and 96.67\% of the time is spent on scanning metadata in ballot filter alone solution, respectively. 


\begin{figure}[t]
	\centering
	\includegraphics[scale=0.34]{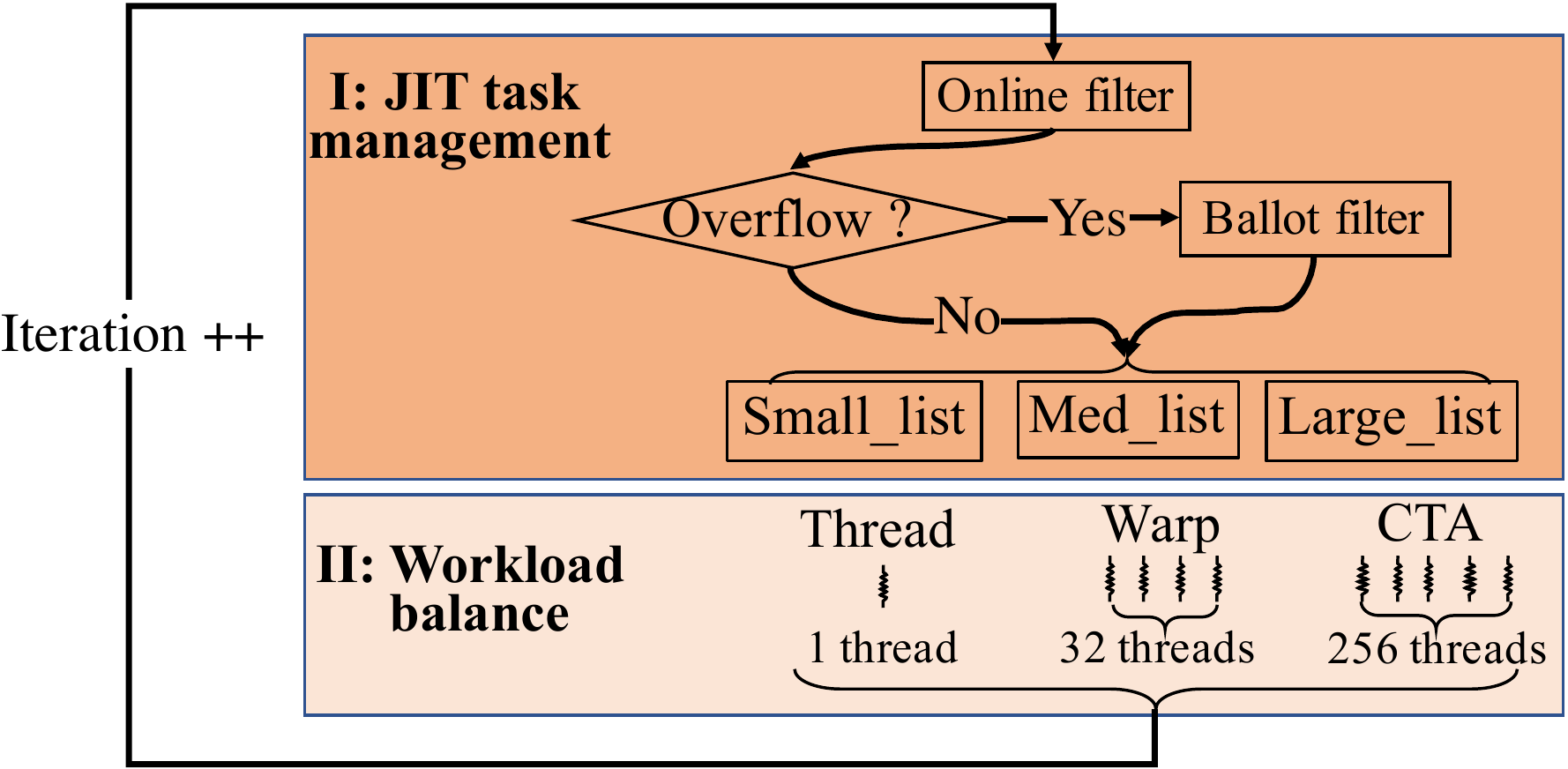}
	\caption{Just-in-time task management. } 
	\label{adaptive-worklist-generation}
\end{figure}

\begin{figure}[t]
	\centering
	\includegraphics[scale=0.3]{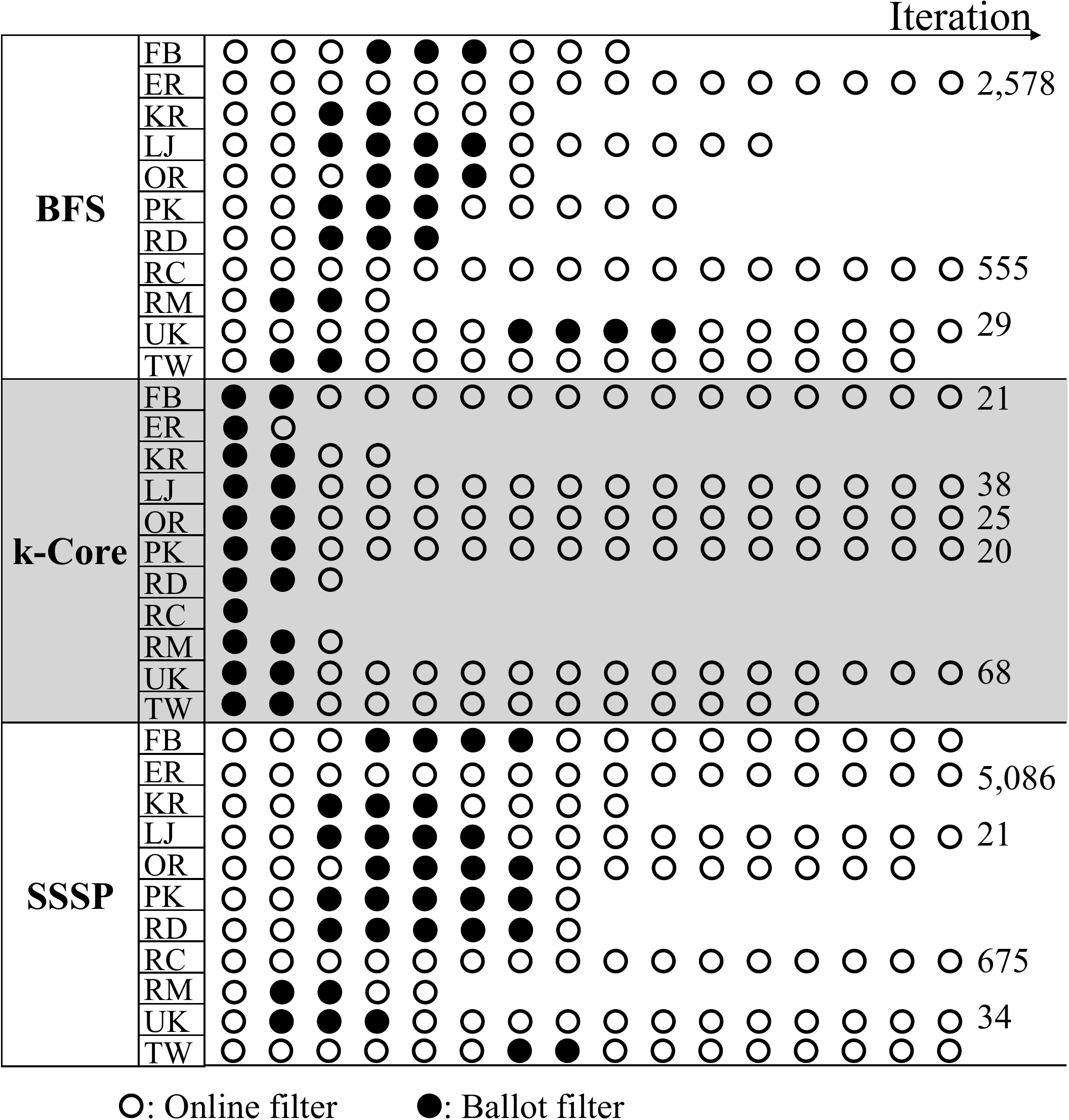}
	\caption{Ballot filter activation patterns. } 
	\label{tech-filter-selection-pattern}
\end{figure}

\vspace{2mm}
\noindent \textbf{Online filter} is designed to accommodate the issue faced by ballot filter. In the first step, this method loads the  active neighbors, updates the destination vertex, and simultaneously records  the active vertices in the thread bin. In the last step, it assembles all thread bins together as the next active list.
When the number of active vertices is small, this approach turns out to be extremely fast.  
Here we use the early stage of SSSP as an example to explain its working process. As shown in  Figure~\ref{tech-filter-example}(c), \{$b$, $d$\} are active vertices, this approach loads their neighbors for computation ({\scriptsize \encircle{c1}}), and immediately records the destination vertices. Eventually, it generates \{$e$, $c$\} as the active list for the next iteration as shown in {\scriptsize \encircle{c2}}. It is also important to note that for online filter, the vertices in the active list may become redundant, and out of order. 

In graph computing, it is possible that one GPU thread may encounter exceeding amount of active vertices, e.g., our tests on Twitter graph shows one GPU thread can reap more than 4,096 active vertices. 
Clearly, one cannot afford such a large thread bin for all threads, thus online filter will inevitably suffer from an overflow problem.
Fortunately, ballot filter largely avoids this issue because it first updates the metadata of active vertices {\scriptsize \encircle{b2}}, which, to some extent, averages out the active vertices across threads in step {\scriptsize \encircle{b3}}.  
Our evaluation also demonstrates such a difference.


\vspace{0.05in}	
\noindent \textbf{Just-In-Time control} adaptively exploits ballot and online filters to retain the best performance. 
As shown in Figure~\ref{adaptive-worklist-generation}, {\simd} always activates the online filter first. 
Once a thread bin overflows, {\simd} will switch on ballot filter to generate the correct task list for the next iteration. 
Interestingly, we find out that various algorithms and graph datasets present different selection patterns which tie closely to the amount of workload, that is, the higher volume of workload often results in the activation of ballot filter.  
As shown in Figure~\ref{tech-filter-selection-pattern}, BFS and SSSP typically use the ballot filter in the middle of the computation and online filter at the beginning and end. 
For high-diameter graphs, BFS and SSSP avoid the use of ballot filter. 
For instance, ER and RC always use the online filter along 2,578, 555, 5,086 and 675 iterations.  k-Core activates the ballot filter at the initial iterations, i.e., typically the first two iterations except RC which only experiences one iteration because all its vertices have $<16$ neighbors. 
At the extreme, BP and PageRank need the ballot filter at exactly the first iteration of computation. 

\begin{figure}[t]
	
	\begin{tabular}{cc}
		\includegraphics[scale=0.34]{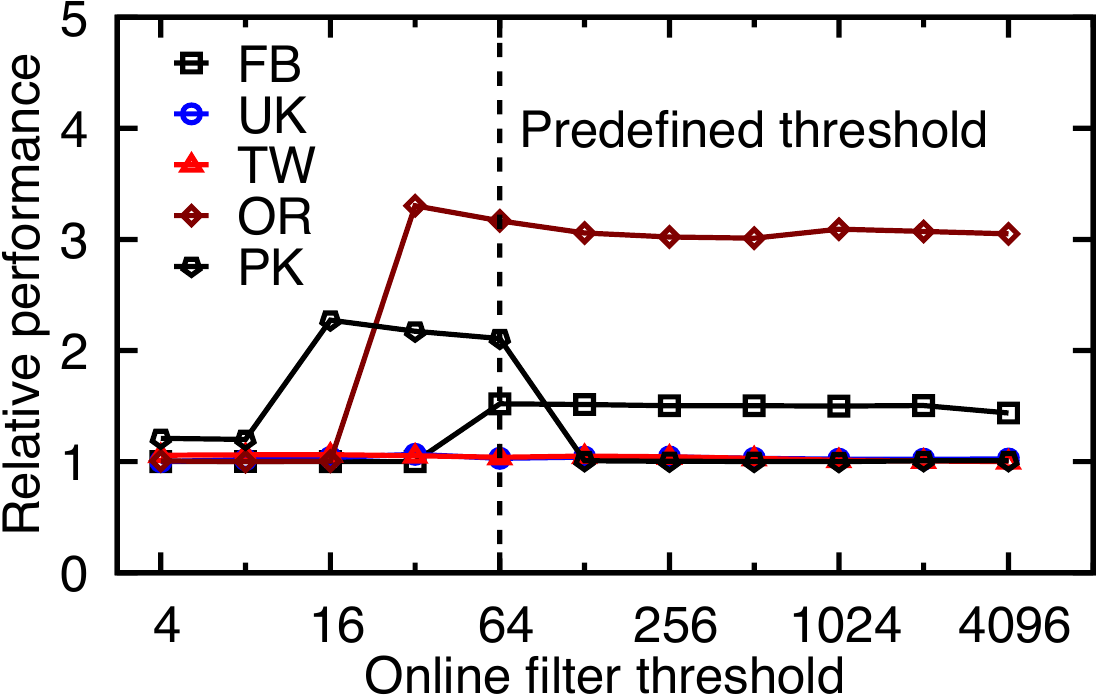}& 
		\hspace{-.1in}\includegraphics[scale=0.34]{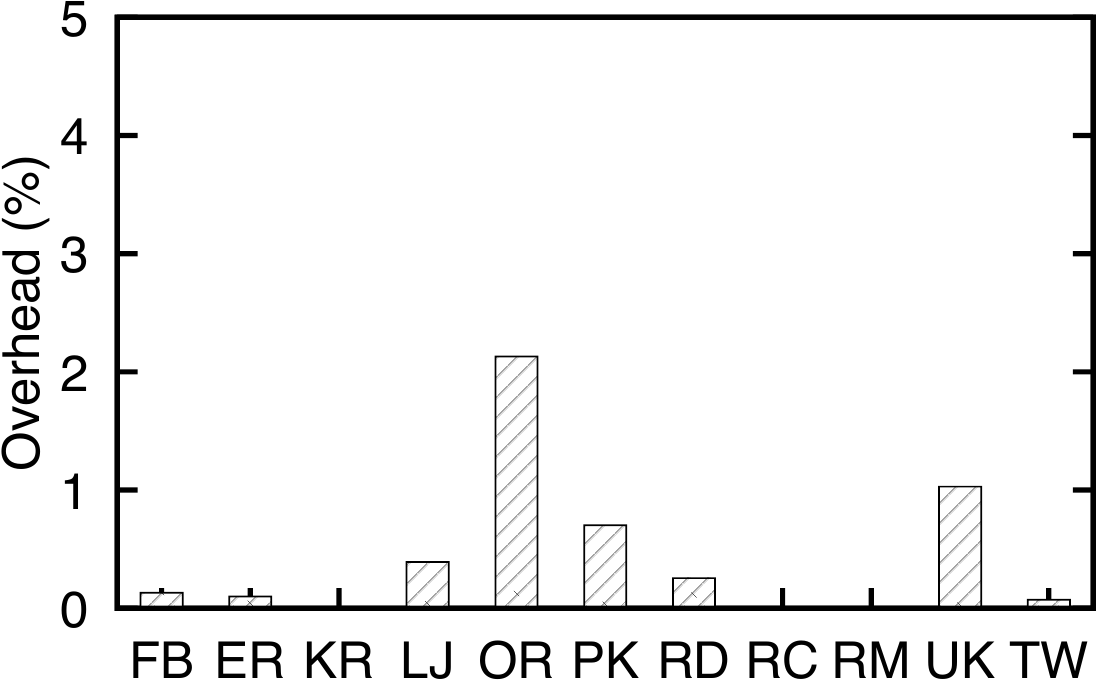}\\
		{\small(a) Relative performance.} & {\small(b) JIT Overhead} \\
	\end{tabular}
	\vspace{-0.1in}
	\caption{The (a) relative performance of JIT management with respective to various online filter overflow thresholds on BFS and (b) the overhead of JIT on SSSP.} 
	\label{exp-online-filter}
	\vspace{-0.2in}
\end{figure}

\vspace{0.05in}
\noindent \textbf{Overflow thresholds for online filter.} Clearly, this parameter directly determines when to switch on ballot filter, thereby affects the overall performance. Figure~\ref{exp-online-filter}(a) presents the normalized performance with respect to various thresholds. As expected, a too low or too high threshold limits the  performance because in either case, {\simd} is forced to switch to ballot filter either too early or too late, leading to performance penalty. As such, in this work we select 64 as the predefined overflow threshold for all algorithms.

\vspace{0.05in}
\noindent
\textbf{Overhead of online filter.}
After switching to ballot filter,  JIT task management also executes the online filter in case it needs to switch back. 
 Figure~\ref{exp-online-filter}(b) studies the overhead of this design. On average, there is 0.02\% slowdown, with the maximum of 2.1\% observed for the OR graph.  
 The reason for the small overhead is because online filter only tracks upto 64 (predefined threshold) active vertices for the next iteration  and this operation is not on the critical path of the execution. 

\vspace{0.05in}
\noindent \textbf{Classification of small, medium and large worklists.} Given GPU thread granularity, we initialize the small, medium and large worklists  to be warp and block sizes (i.e., 32 and 128), respectively. Our further investigation shows, for the separator of small to medium worklists, the performance stays stable in the range of [4, 128], and for medium and large worklists [128, 2048]. 
We find the performance starts to drop beyond these ranges.

\section{Push-Pull Based Kernel Fusion}
\label{sec-fuse}


Kernel fusion~\cite{wahib2014scalable},  a common optimization for a collection of iterative GPU applications, such as graph computing and deep learning~\cite{abadi2016tensorflow,paszke2017automatic,jia2014caffe,chetlur2014cudnn,chen2015mxnet}, reduces expensive overhead of kernel invocation, as well as minimizes the global memory traffic as the life time of registers and shared memory is limited in each kernel. {However, traditional efforts, such as Gunrock~\cite{wang2015gunrock} and Xiao et al~\cite{xiao2010inter}, fail to achieve cross the global barrier kernel fusion.
This section starts with our observation and analysis of potential deadlock in the mainstream global barrier design~\cite{xiao2010inter,yan2013streamscan} and subsequently introduces a light-weighted deadlock free solution which enables the global thread synchronization within the fused kernel. However, aggressive kernel fusion requires a large amount of the registers and thus supports fewer parallel warps which could hurt the overall performance. 
 To this end, we introduce a push-pull based kernel fusion strategy to minimize both the kernel invocation times and register consumption.}


\vspace{2mm}
\noindent \textbf{Software global barrier} is needed to enable the balanced kernel fusion.
Generally speaking, this approach uses an array -- $lock$ -- to synchronize all GPU threads upon arrival and departure. 
During the processing, it assumes the thread CTA as the monitor while the remaining threads as workers. 
At arrival, each worker CTA updates its own $status$ in $lock$.
Once all worker CTAs have arrived, the monitor changes the $statuses$ of all CTAs to \textit{departure}, allowing all threads to proceed forward. 

\begin{figure}[t]
	\centering
	\includegraphics[scale=0.34]{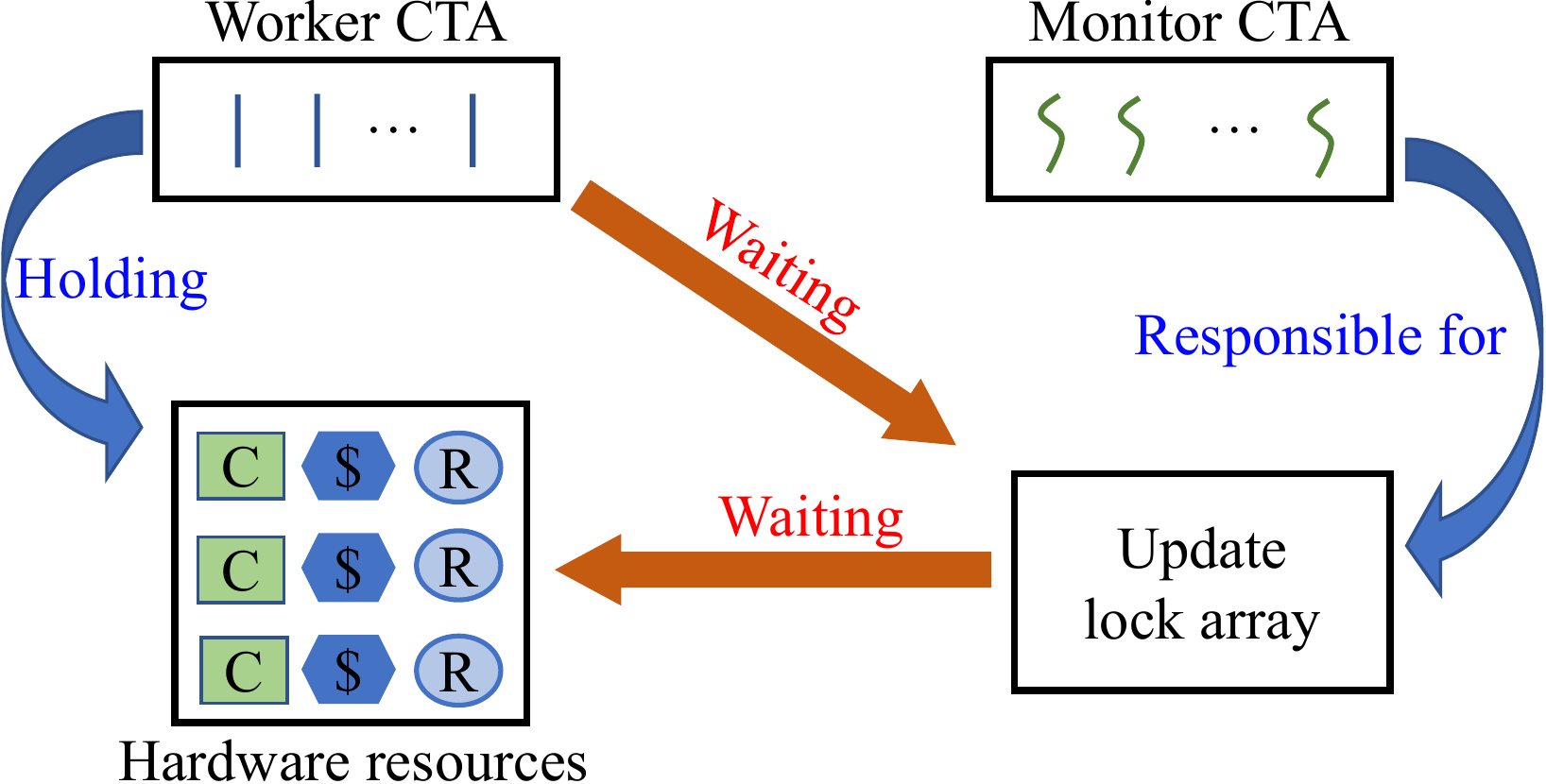}
	\caption{Deadlock problem in software global barrier, where `C', '\$', and `R' represent CUDA core, L1 cache and register, respectively.} 
	\label{sync-challenge}
		\vspace{-0.1in}
\end{figure}

This approach, unfortunately, suffers from potential deadlock~\cite{xiao2010inter}, as illustrated in  
Figure~\ref{sync-challenge}. 
Specifically, the worker thread CTAs may hold all GPU hardware resources, such as streaming processors, registers and shared memory, while waiting for the monitor to update the $lock$ array. 
In the meantime, the monitor cannot update the $lock$ array, due to lack of hardware resources (e.g., thread over subscription).

\noindent 
\textbf{Compiler-based deadlock free barrier}. 
{\simd} utilizes the barrier in a way to ensure that every CTA, regardless of a work or the monitor, can obtain hardware resources when needed. 
This is achieved through comparing the resources needed by the kernels, against the total available resources.  
Based on the GPU architecture, we can  obtain the total amount of registers ($\#registerPerSMX$) that can be provided by each streaming processor, e.g., 65,536 registers of NVIDIA K40 GPUs and 32,768  from K20 GPUs. 
On the other hand, we can collect the register consumption ($\#registerPerThread$) of each kernel at the compilation stage.
Putting these numbers together, {\simd} is able to calculate the appropriate thread configuration for the kernels. 

\begin{table*}[t]
	\centering
	\scriptsize
	\caption{Register consumption for various kernels.}
	\label{table-compile}
	\begin{tabular}{|c|c|c|c|c|c|c|c|c|c|c|c|}
		\hline
		\multirow{2}{*}{Kernel} & \multicolumn{4}{c|}{Push (no fusion)} & \multicolumn{4}{c|}{Pull (no fusion)} & \multicolumn{2}{c|}{Selective fusion} & \multirow{2}{*}{All fusion} \\ \cline{2-11}
		& Thread & Warp & CTA & Task mgt & Thread & Warp & CTA & Task mgt & push & pull &  \\ \hline
		Register consumption & 26 & 27 & 28 & 24 & 24 & 24 & 22 & 30 & 48 & 50 & 110 \\ \hline
		Kernel launching count & \multicolumn{8}{c|}{up to 40,688} & \multicolumn{2}{c|}{3} & 1 \\ \hline
	\end{tabular}
\end{table*}

The number of CTA can be computed as follows:
{\small
	\begin{equation}\label{eq-thread-conf}
	\#CTA = floor(\frac{\#registersPerSMX}{\#registersPerThread\cdot\#threadsPerCTA})\cdot\#SMX
	\end{equation}
}
where $\#threadsPerCTA$ is configured by a user, i.e., 128 by default.
For example, when deploying a kernel, each thread consumes 110 registers, and on K40 that contains 15 SMXs, each of which contains 65,536 registers. 
If $\#threadsPerCTA$ is set to 128, one gets $\#CTA = ceil(\frac{65536}{110\times 128})\times 15 = 60$. 
As a result, we can configure this kernel as CTA and thread count per CTA as 60 and 128, respectively. 

\change{
Notably, portable Inter-Block Barrier ~\cite{sorensen2016portable} is closely relevant to our effort. However, this method proposes extremely complicated thread block management mechanism that requires to distinguish whether one thread block will execute useful workloads or not during runtime. This requires  nontrivial programmer efforts and scheduling overhead. In comparison, our method achieves this deadlock-free configuration before runtime and is completely transparent to the end users.
}

\vspace{.04in}
\noindent
\textbf{Push-Pull based kernel fusion}. 
As shown in  Table~\ref{table-compile}, the register consumption (using the compilation flag \textit{-Xptxas -v})  increases from average 25 to 110, that is 4.4$\times$ before and after kernel fusion.
It becomes clear that we need a balanced fusion strategy that keeps both register consumption and kernel invocation low.
To this end, {\simd} leverages the push-pull model used in the graph algorithms.
That is, such algorithms often use push or pull based computing in several consecutive iterations.
For example, BFS and SSSP utilize push in the first and last iterations, and pull in between. 
In contast, k-Core conducts pull at the beginning while push in the end. 

The idea of push-pull based kernel fusion is to fuse kernels around the pull and push computing.
In other words, for the push-based iterations, {\simd} fuses different compute kernels (for thread, warp, CTA), as well as task management kernel, into one push kernel.
The kernel only terminates when the computation finishes or it needs to switch to pull computing according to the criterion discussed in Section~\ref{sec-api}.
Similar optimizations are done for the pull-based iterations. 

\begin{figure*}
\begin{center}
		\subfloat[{ BFS}]{\includegraphics[scale=0.36]{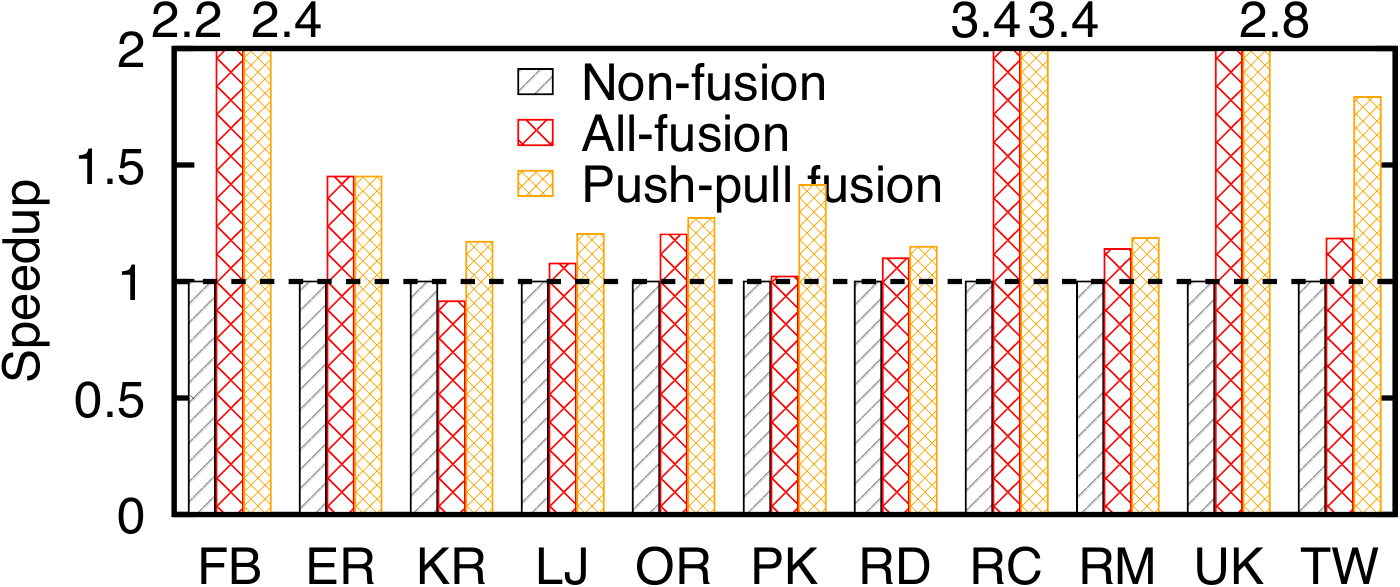}} \quad 
		\subfloat[{ BP}]{\includegraphics[scale=0.36]{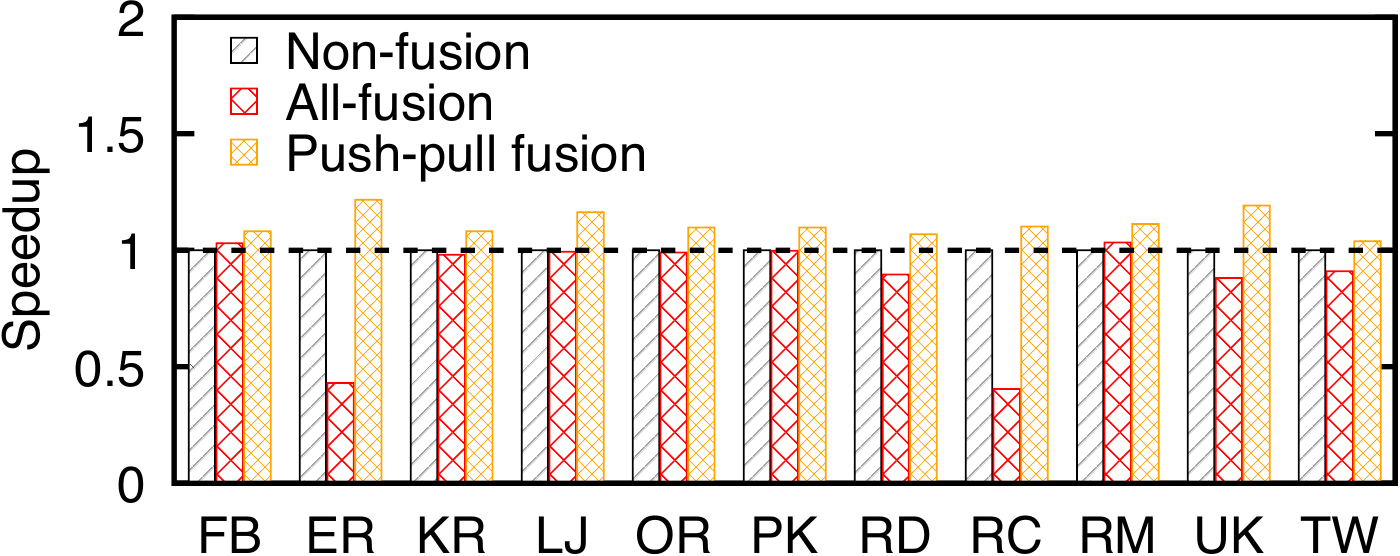}} \quad
		\subfloat[{ k-Core}]{\includegraphics[scale=0.36]{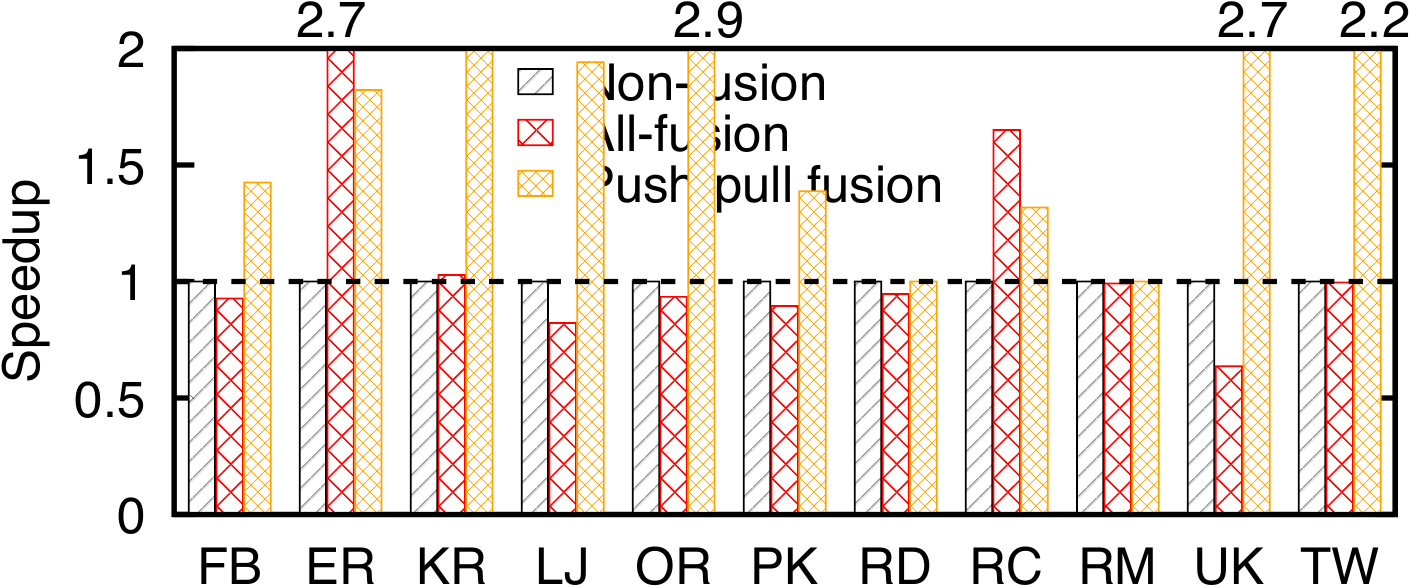}} \\
		\hspace{-.3in}\subfloat[{ PageRank}]{\includegraphics[scale=0.36]{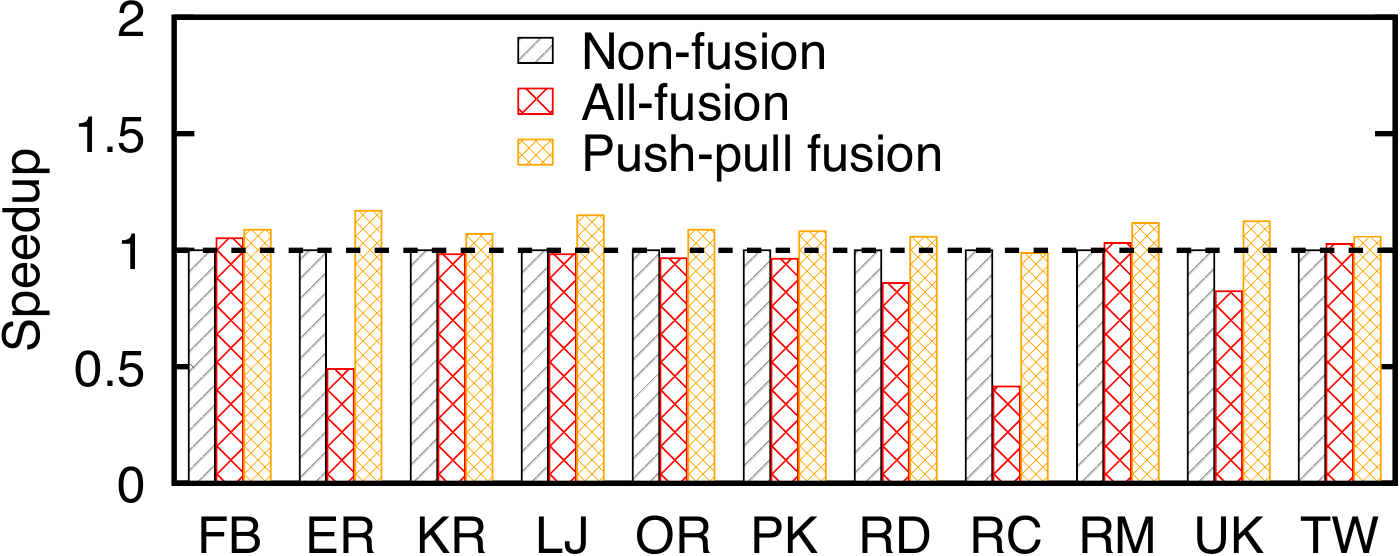}} \quad
		\subfloat[{ SSSP}]{\includegraphics[scale=0.36]{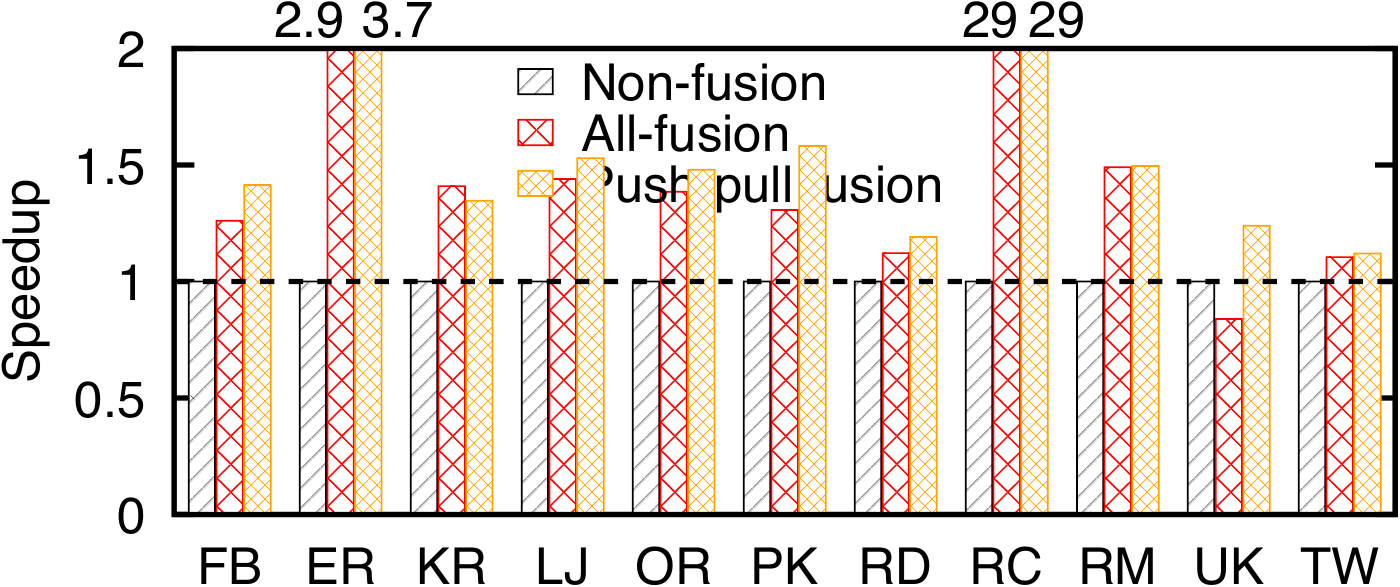}}
\end{center}
	\vspace{-.3in}
	\caption{ Benefit of push-pull based kernel fusion, normalized to the performance of no fusion. }
	
	\label{exp-selective-fusion}
	\vspace{-.2in}
\end{figure*}

Using the new push-pull based fusion, the register consumption decreases to 48 and 55 thus increases the configurable thread count by 50\%. Together, our evaluations demonstrate a 25\% performance improvement in Figure~\ref{exp-selective-fusion} of Section~\ref{eval-tech}.

\begin{figure}[t]
	\centering
	\includegraphics[scale=0.25]{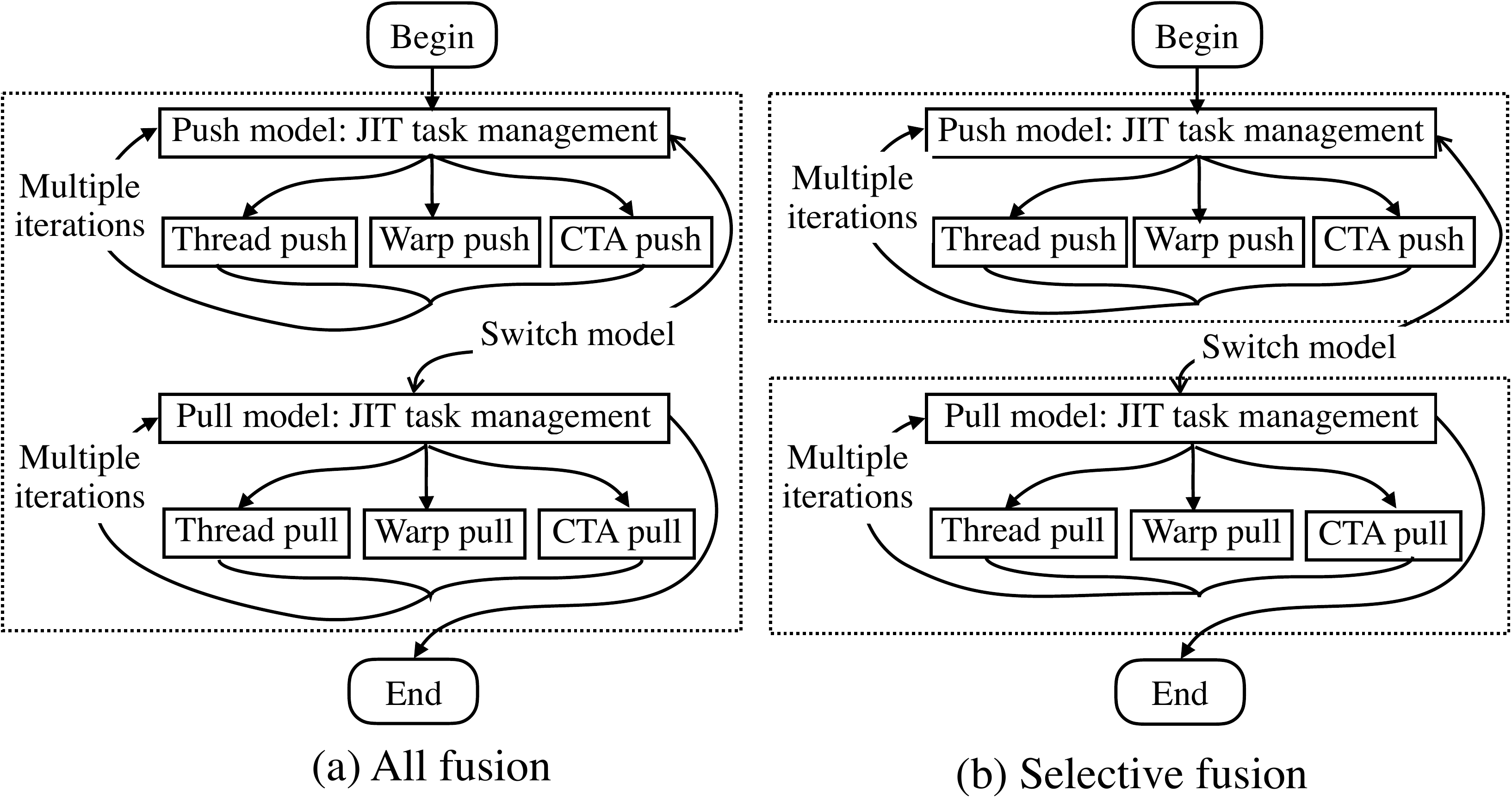}
		\vspace{-0.1in}
	\caption{Consecutive iterations from graph algorithms often cluster to push and model computation separately: (a) all fusion, (b) selective fusion.} 
	\label{tech-select-fuse}
		\vspace{-0.3in}
\end{figure}

Table~\ref{table-compile} presents the register consumption  and kernel invocation of different kernel fusion techniques. 
By using the push-pull based kernel fusion, the kernel relaunch is reduced to 3 while its register consumption is cut by half. 
In Figure~\ref{exp-selective-fusion}, we will later show that this technique brings upto 80\% performance benefit.

\section{Graph Algorithms and Datasets}
\label{sec-app}

In addition to  SSSP that is discussed in Section~\ref{sec-api}, this section further presents a variety of algorithms which are implemented on {\simd} to examine the expressiveness of ACC programming model, and performance impacts of task management and  kernel fusion techniques.

\vspace{.05in}
\noindent
\textbf{BFS}~\cite{hang2015enterprise} traverses a graph level by level. At each level, it loads all neighbors that are connected to vertices visited in the preceding level, inspects their statuses (metadata), and subsequently marks those unvisited neighbors as active for the next iteration. 
Notably, BFS conducts synchronizations at the end of each level, relies on vote to combine the updates. 
During the entire process of traversal, BFS typically experiences light workload at the beginning and end of the computation while heavy workload in the middle. 

\vspace{.05in}
\noindent
\textbf{Belief propagation} (BP), also known as sum-product message passing algorithm, infers the posterior probability of each event based on the likelihoods and prior probabilities of all related events. Once modeled as a graph (Bayesian network or Markov random fields), each event becomes a vertex with all incoming vertices and edges as related events and corresponding likelihoods. In BP, vertex possibility is the metadata. 

 
\vspace{.05in}
\noindent
\textbf{$k$-Core} (KC), which is widely used in graph visualization application~\cite{liu2017graphene,montresor2013distributed}, iteratively deletes the vertices whose degree is less than $k$ until all remaining vertices in this graph possess more than $k$ neighbors. 
$k$-Core experiences large volume of workloads at initial iterations and follows with light workloads. This work uses a default value of $k=16$.

\vspace{.05in}
\noindent
\textbf{PageRank} (PR)~\cite{page1999pagerank} updates the rank value of one vertex based on the contribution of all in-neighbors iteratively till all vertices have stable rank values. Because the contributions of in neighbors are summarized to the destination vertex, we start PageRank with the pull model and $agg\_sum$ as the merge operation. At the end of PageRank, we switch to the push model because the majority of the vertices are stable~\cite{zhang2014maiter}. The switch is decided by a decision tree.

\vspace{.05in}
\noindent
\textbf{Storage Format}. {\simd} employs \textit{compressed sparse row} (CSR) format to store the graph. For undirected graph, we only need to store the out-neighbors of each vertex. For directed graph, we store both out-neighbors and in-neighbors of each vertex to support the push and pull based processing.

\begin{table}[t]
	\centering
	{
		\small
		\caption{Graph Dataset}
		\label{tab-graph}
		\begin{tabular}{l l r r r r|r|}
			\hline			
			Graph Name  & Abbrev. & Vertex Count & Edge Count      \\
			\hline
			\hline
			\\[-.9em]
			Facebook    & FB      & 16,777,215   & 775,824,943      \\
			Europe-osm  & ER      & 50,912,018   & 108,109,319 \\
			Kron24      & KR      & 16,777,216   & 536,870,911       \\
			LiveJournal & LJ      & 4,847,571    & 136,950,781       \\
			Orkut       & OR      & 3,072,626    & 234,370,165      \\
			Pokec       & PK      & 1,632,803    & 61,245,127       \\
			Random      & RD      & 4,000,000    & 511,999,999     \\
			RoadCA-net  & RC      & 1,971,281    & 5,533,213   \\
			R-MAT       & RM      & 3,999,983    & 511,999,999    \\
			UK-2002     & UK      & 18,520,343   & 596,227,523      \\
			Twitter     & TW      & 25,165,811   & 787,169,139\\
			\hline		
		\end{tabular}
	}
\end{table}

\vspace{.05in}
\noindent
\textbf{Graph Benchmarks}. We evaluate on a wide range of graphs as shown in Table~\ref{tab-graph}, which falls into four types, i.e., social networks, road maps, hyperlink web and synthetic graphs. Particularly, Facebook~\cite{gjoka2011practical}, LiveJournal~\cite{snap}, Orkut~\cite{snap}, Pokec~\cite{snap}, and Twitter~\cite{kwak2010twitter} are common social networks. Europe-osm~\cite{europe_osm} and RoadCA-net~\cite{matrix_market} are two large roadmap graphs, and UK-2002~\cite{matrix_market} is a web graph. 
Furthermore, we use Graph500 generator to generate Kron24~\cite{chakrabarti2004r}, and  GTgraph~\cite{gt_graph} for R-MAT and random graphs. 
Europe-osm and RoadCA-net are high diameter graphs, with 2570 and 555 as their diameters, respectively. LiveJournal, Pokec, Twitter and UK-2002 are medium diameter graphs, i.e., 10 - 30 as their diameters. The diameters of the remaining graphs are all smaller than 10.
For graphs without edge weight, we use a random generator to generate one weight for each edge similar to Gunrock~\cite{wang2015gunrock}.

\newcommand{\mcol}{0.35in}
\begin{table*}[h]
	\centering
\scriptsize
	\caption{Runtime (ms) of {\simd} and Gunrock, and Galois. A K40 GPU is used to test {\simd} and Gunrock, and a CPU with 28 threads for Galois. The \textbf{blank space} indicates the test cannot complete for the given algorithm and graph.}
	\label{exp-comp-soa}
		\begin{tabular}{lllllllllllllc}
		\hline
Alg & System & FB & ER & KR & LJ & OR & PK & RD & RC & RM & UK & TW & Avg. speedup \\ \hline
& {\simd} & \cellcolor[HTML]{67FD9A}198 & \cellcolor[HTML]{67FD9A}400 & \cellcolor[HTML]{67FD9A}130 & \cellcolor[HTML]{67FD9A}59 & \cellcolor[HTML]{67FD9A}40 & \cellcolor[HTML]{67FD9A}20 & 82 & \cellcolor[HTML]{67FD9A}15 & 47 & 308 & \cellcolor[HTML]{67FD9A}210 & - \\
& CuSha &  &  & 988 & 224 & 341 & 72 & 435 & 297 & 674 & 4298 &  & 9.6 \\
& Gunrock & 685 & 849 & 677 & 71 & 225 & 44 & 647 & 146 & 506 & 312 & 697 & 4.8 \\
& Galois & 482 & 1068 & 140 & 139 & 42 & 34 & \cellcolor[HTML]{67FD9A}48 & 53 & 65 &\cellcolor[HTML]{67FD9A} 229 & 322 & 2.1 \\
\multirow{-5}{*}{BFS} & Ligra & 1086 & 1426 & 176 & 89 & 51 & 31 & 88 & 48 & \cellcolor[HTML]{67FD9A}40 &  & 496 & 2.4 \\ \hline
& {\simd} & \cellcolor[HTML]{67FD9A}1553 & \cellcolor[HTML]{67FD9A}346 & \cellcolor[HTML]{67FD9A}1141 & 236 & 435 & \cellcolor[HTML]{67FD9A}118 & \cellcolor[HTML]{67FD9A}1105 & \cellcolor[HTML]{67FD9A}13 & \cellcolor[HTML]{67FD9A}800 & \cellcolor[HTML]{67FD9A}637 & \cellcolor[HTML]{67FD9A}1525 & - \\
& CuSha &  &  & 1704 & \cellcolor[HTML]{67FD9A}182 & \cellcolor[HTML]{67FD9A}323 & 180 & 1402 & 15 & 886 &  &  & 1.2 \\
& Gunrock & 3004 & 884 & 3129 & 275 & 927 & 166 & 2963 & 43 & 2208 & 784 & 3180 & 2.1 \\
& Galois & 4552 & 603 & 3069 & 424 & 1061 & 218 & 3576 & 20 & 2067 & 842 & 4178 & 2.3 \\
\multirow{-5}{*}{PR} & Ligra & 16780 & 1368 & 2000 & 1324 & 1786 & 310 & 809 & 35 & 1703 &  & 9360 & 4 \\ \hline
& {\simd} & \cellcolor[HTML]{67FD9A}1816 & \cellcolor[HTML]{67FD9A}1080 & \cellcolor[HTML]{67FD9A}998 & \cellcolor[HTML]{67FD9A}284 & \cellcolor[HTML]{67FD9A}604 & \cellcolor[HTML]{67FD9A}143 & 1505 & \cellcolor[HTML]{67FD9A}223 & \cellcolor[HTML]{67FD9A}478 & \cellcolor[HTML]{67FD9A}703& \cellcolor[HTML]{67FD9A}1344 & - \\
& CuSha &  & 519674 & 1663 & 692 & 1120 & 260 & 1610 & 438 & 1236 &  &  & 62 \\
& Gunrock &  & 1206 & 1220 & 431 & 1259 & 336 & 5059 & 229 &  &  &  & 1.8 \\
& Galois & 161596 &  & 8485 & 1785 & 1166 & 356 & \cellcolor[HTML]{67FD9A}747 & 3440 & 5877 & 9081 &1818 & 15 \\
\multirow{-5}{*}{SSSP} & Ligra & 14067 & 3043 & 2893 & 1627 & 1567 & 605 & 3353 & 301 & 2783 & 1300 & 5217 & 3.7 \\ \hline
& {\simd} &\cellcolor[HTML]{67FD9A}366&\cellcolor[HTML]{67FD9A}	78&\cellcolor[HTML]{67FD9A}	131&\cellcolor[HTML]{67FD9A}	60&\cellcolor[HTML]{67FD9A}	63&\cellcolor[HTML]{67FD9A}	33&\cellcolor[HTML]{67FD9A}	10&\cellcolor[HTML]{67FD9A}	4&\cellcolor[HTML]{67FD9A}	19&\cellcolor[HTML]{67FD9A}	151&\cellcolor[HTML]{67FD9A}	277 & -\\
\multirow{-2}{*}{\textit{k}-Core} & Ligra & 6337&	1167&	2813&	1707&	1700&	654&	27&	36&	235&	6627&	5783 & 20\\\hline
	\end{tabular}
\vspace{-.2in}
\end{table*}

\section{Experiments}
\label{sec-exp}

We implement {\simd}\footnote{{\simd} will be released in open source upon the paper publication.} with 5,660 lines of CUDA and C++ code. 
All the algorithms presented in Section~\ref{sec-app} are implemented with around 100 lines of C++ code. 
The source code is compiled by GCC 4.8.5 and NVIDIA nvcc 7.5 with the optimization flag as O3. In this work, we evaluate {\simd} on a Linux workstation with two Intel Xeon E5-2683 CPUs (14 physical cores with 28 hyperthreads), and 512GB main memory.
Throughout the evaluation, we use uint32 as the vertex ID and uint64 as index and evaluate our system on NVIDIA K40 GPUs unless otherwise is specified. We also test {\simd} on earlier K20 and latest P100  GPUs. 
The timing is started once the graph data is loaded in GPU global memory. Each result is reported with an average of 64 runs. 

\subsection{Comparison with State-of-the-art}

Table~\ref{exp-comp-soa} summarizes the runtime of {\simd} against Galois and Gunrock which are state-of-the-art CPU and GPU graph processing systems, respectively, as well as CuSha (GPU) and Ligra (CPU), two popular graph frameworks.  The take aways of this table are two folds. 

First, {\simd} is both space efficient and robust. As one can see, since CuSha requires edge list as the input for computation, it cannot accommodate large graphs (e.g., FB and TW) across all algorithms. Besides, since Gunrock requires large amount of space for batch filter, it suffers out of memory (OOM) error for all larger graphs in SSSP. Even CPU systems (Galois and Ligra) enjoys affluent memory space (512 GB) from CPU, they cannot converge to a result for high diameter graphs. That is, Galois cannot converge for SSSP on ER while Ligra fails to obtain result for BFS on UK graph.

Second, {\simd} outperforms all graph processing frameworks. 
In general, {\simd} is 24$\times$,  2.9$\times$, 6.5$\times$ and 3.3$\times$ faster than CuSha, Gunrock, Galois and Ligra, respectively.
In BFS, {\simd} bests CuSha, Gunrock, Galois and Ligra by 9.6$\times$, 4.8$\times$, 2.1$\times$ and 2.4$\times$, respectively. 
We also notice that {\simd} is slower than Galois on the RD graph because workload balancing brings negligible benefits to uniform-degree graph (RD). Also, {\simd} is slightly worse than Ligra on RM graph since this graph only has a diameter of four thus both the optimization of JIT task management and kernel fusion brings trivial benefits to GPU based graph systems, as evident by the much lower performance on CuSha and Gunrock.

In PageRank, {\simd} achieves 1.2$\times$, 2.1$\times$, 2.3$\times$ and 4$\times$ speedups over CuSha, Gunrock, Galois and Ligra, respectively. 
Note, even CuSha cannot support all large graphs due to large memory space consumption, it performs roughly similar to {\simd} with even outperforming {\simd} on LJ and OR. This is generally because PageRank tends to be computation intensive and needs to compute all edges, curbing the benefits of task management and kernel fusion.
However, edge list format (of CuSha) doubles the memory consumption, facing the crisis of OOM for all large graphs. 
 
For SSSP, {\simd} wins 21$\times$, on average, over all four projects. We project {\simd} to better outperform all systems than observed for BFS algorithm because SSSP experiences more iterations with larger volume of active tasks, placing more favor towards ACC model, JIT task management and push-pull based kernel fusion. However, because Gunrock fails to accommodate all large graphs, our benefits cannot surface -- ending with merely 1.8$\times$ speedup. Second, CuSha spends 519,674 ms on the high diameter ER graph which is 480$\times$ slower than {\simd} because task management is absent from CuSha. We also notice Galois performs better than {\simd} in RD, again, due to the uniform degree distribution phenomenon.

\begin{figure*}[t]
	
	\begin{tabular}{ccc}
		\includegraphics[scale=0.36]{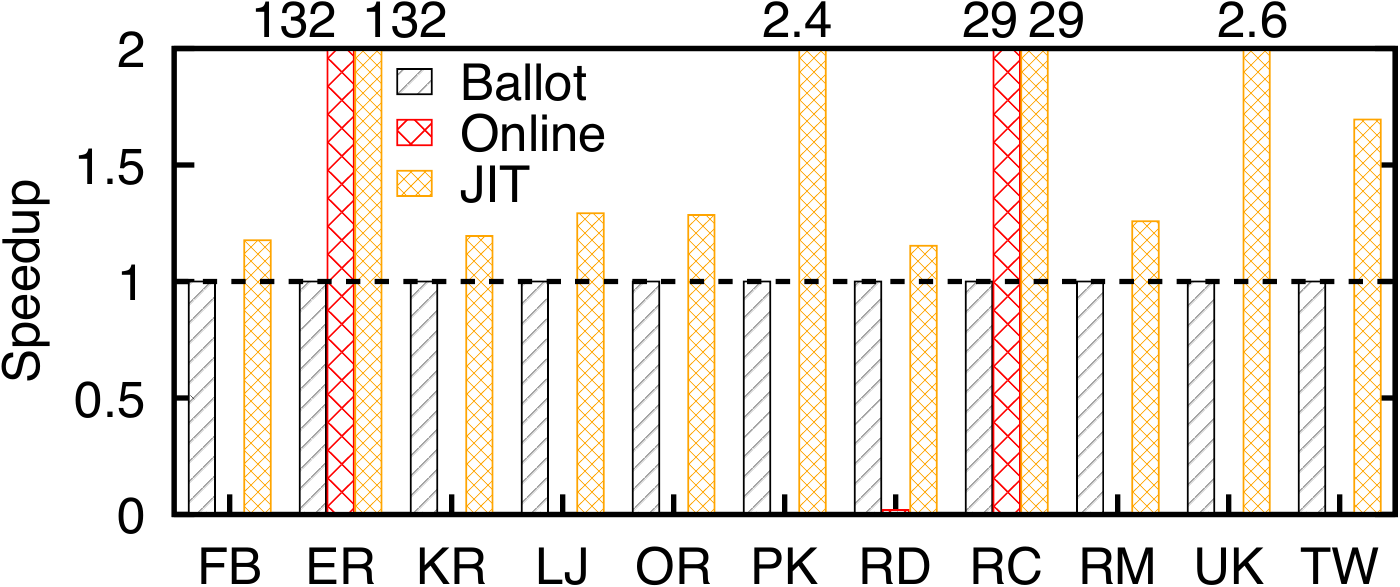}& 		\includegraphics[scale=0.36]{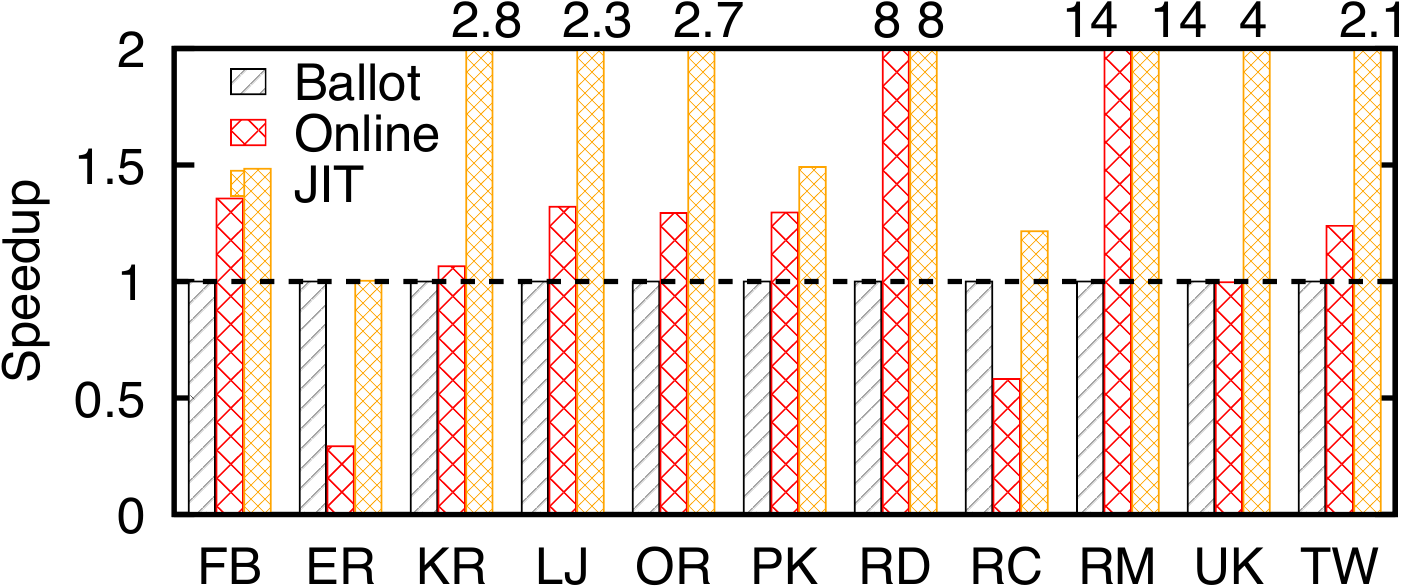}& 
		\includegraphics[scale=0.36]{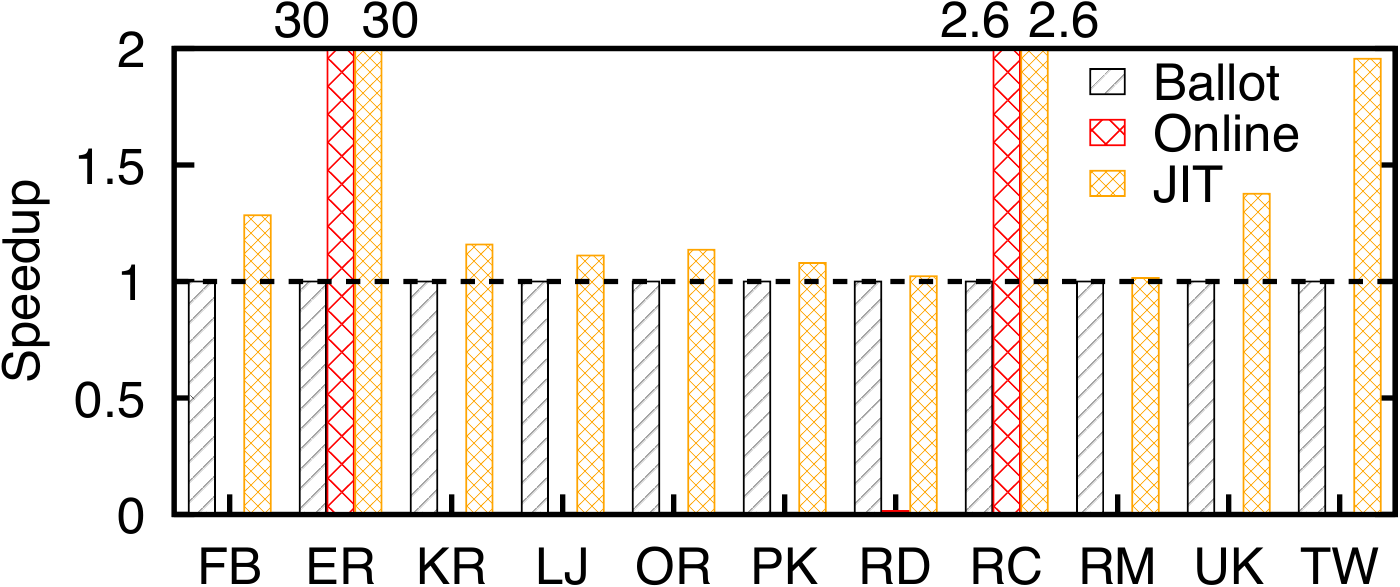}\\
		{\small(a) BFS} & {\small(b) k-Core} & {\small(c) SSSP} \\
	\end{tabular}
	\vspace{-0.1in}
	\caption{Benefit of just-in-time task management, normalized to the performance of the ballot filter.} 
	\label{exp-task-manage}
		\vspace{-0.1in}
\end{figure*}

\change{
For \textit{k}-Core, where $k$ = 32, {\simd} wins Ligra by 20$\times$. Such a striking advantage comes from three parts. First, as reflected by Figure~\ref{exp-task-manage}(b), $k$-Core generates extensive amount of workload variations thus benefits tremendously from JIT task management. Second, $k$-Core's iterative nature also enjoys the benefits from push-pull based kernel fusion, as shown in Figure~\ref{exp-selective-fusion}(c). Lastly, the flexibility of ACC allows innovative $k$-Core algorithm designs -- we will stop further subtracting the degree of destination vertex once the destination vertex's degree goes below $k$ -- this reduces tremendous unnecessary updates. 
Note comparisons of Belief Propagation, as well as other systems for $k$-Core are not included because those systems fail to support such algorithms. 
}

\subsection{Benefits of Various Techniques}
\label{eval-tech}

This section studies the performance impacts brought by JIT task management and push-pull based kernel fusion. 
As we have presented in Section~\ref{sec-sched}, JIT task management only works for applications that experience workload variations, that is, BFS, k-Core and SSSP. 
On the other hand, push-pull based kernel fusion is applicable for all five algorithms

On average, JIT task management presented in Figure~\ref{exp-task-manage}, is 16$\times$, 26$\times$ and 4.5$\times$ faster than the ballot filter for BFS, k-Core and SSSP. 
As expected, online filter alone cannot work for many graphs, particularly large ones, e.g., Facebook, Twitter and UK2002 graphs in BFS and SSSP. 
Without considering overflow problem (ER and RC graphs), JIT task management adds a small 1-2\% overhead on top of the online filter on BFS and SSSP. 

On k-Core, JIT task management is, on average, 28.5$\times$ and 5\% faster than ballot and online filter, respectively.  
We also observe that the ballot filter outperforms the online filter on ER and RC graphs by 3.4$\times$ and 1.7$\times$, because  k-Core removes a large volume of vertices which favors the former to produce a non-redundant and sorted work list.

Push-pull based kernel fusion brings, on average, 43\% and 25\% improvement over non-fusion and all-fusion across all algorithms and graphs. 
In particular, push-pull based kernel fusion tops non-fusion by  74\%, 11\%, 85\%, 10\% and 66\% on BFS, BP, k-Core, PageRank and SSSP. 
BFS, k-Core and SSSP achieves more performance gains because they are not computation intensive and tend to run a higher number of iterations. For all fusion, our new kernel fusion is 55\%, 6\%, 62\%, 25\% and 11\% faster on BFS, BP, k-Core, PageRank and SSSP. 
It is important to note that all fusion is not always beneficial, i.e., all fuse option of PageRank is average 13\% slower than no fusion because all fusion limits the amount of configurable threads. However, for memory intensive applications, like BFS and SSSP on ER and RC, all fusion is on average 2$\times$ better.

\subsection{Performance on Different GPUs}

We also evaluate {\simd}, Gunrock and CuSha on various GPU models, such as K20 and P100 GPUs. It is not surprising to see tht {\simd} presents the biggest performance gain on the latest GPUs. In detail, {\simd} on K40 and P100 performs 1.7$\times$ and 5.1$\times$ better than K20 GPU. In contrast, Gunrock merely gets 1.1$\times$ and 1.7$\times$ performance improvement when moving from K20 to K40 and P100, respectively. Similarly for CuSha, its performance on K40 and P100 are 1.2$\times$ and 3.5$\times$ better than K20, respectively.  The root cause of this disparity is that  {\simd}'s kernel fusion technique can dynamically configure its GPU kernels to the fitting thread count on the corresponding hardware so as to achieve the peak performance. For instance, the thread count increases by 1.2$\times$ and 5.1$\times$ on K40 and P100 than on K20 for BFS.
\section{Related Work}
\label{sec-related}

%


Recent advance in graph computing falls in algorithm innovation~\cite{meyer1998delta,zhang2014maiter}, framework  developments~\cite{malewicz2010pregel,gonzalez2012powergraph,low2010graphlab,kyrola2012graphchi,liu2017graphene,zheng2015flashgraph,zhu2015gridgraph,han2013turbograph,shun2013ligra,shao2013trinity,roy2013x,hong2012green,nguyen2013lightweight,roy2015chaos,wang2015graphq,chen2014computation,xie2015sync,zhang2015numa,shi16fast,zhang2017sub,wang2018fast,nodehi2018tigr,zhang2017making,zhang2018wonderland,zhang2016exploring,ai2017squeezing,wu2015g,miao2015immortalgraph,han2014chronos,cheng2012kineograph} and accelerator optimizations~\cite{wang2015gunrock,hang2015enterprise,merrill2012scalable,khorasani2014cusha,zhang2018graphp,prabhakaran2012managing}. This section covers relevant work from three aspects: programming model, task management and kernel fusion. 

Besides edge and vertex centric models, there are also other models that make various trade-offs between simplicity and performance. For instance, ``think like a graph"~\cite{tian2013think} requires each vertex to obtain the view of the entire partition on one machine in order to minimize the communication cost. 
Furthermore, domain specific programming language systems, such as Galois~\cite{nguyen2013lightweight}, Green-Marl~\cite{hong2012green} and Trinity~\cite{shao2013trinity}, allow programmers to write single-threaded source code while enjoying multi-threaded processing. 
In comparison, {\simd} decouples the goal of programming simplicity and performance: with ACC, {\simd} ultimately designs a data-parallel abstraction for deploying irregular graph applications on GPU. With JIT task management and push-pull based kernel fusion, {\simd} pushes the performance towards a magnitude faster than state-of-the-art CPU and GPU frameworks.



Task management is an important optimization for GPU-based graph computing. Besides batch filter~\cite{wang2015gunrock,merrill2012scalable}, there also exist other task management approaches -- strided filter~\cite{hang2015enterprise,liu2016ibfs} and atomic filter~\cite{luo2010effective}. Particularly, strided filter resembles ballot filter but the former one experiences strided memory access when scanning the metadata thus performs up to 16$\times$ worse than ballot filter. Atomic filter relies is similar to online filter but it relies on atomic operation to put active vertices into global active list which suffers from orders of magnitude slow down than online filter. Besides ballot and online filter bests batch, stride and atomic filter, {\simd} goes further via introducing a JIT controller to adaptively use online filter and ballot filter to further improve the performance. We also find that JIT task management can be exploited to help manage active lists for other applications such as warp segmentation~\cite{khorasani2015scalable} and CSR5~\cite{liu2015csr5}.

Kernel fusion affects applications far beyond graph computations. {\simd} demonstrates its benefits in graph computing and Belief Propagation (BP) applications. {\simd} is closely related to global software barrier~\cite{xiao2010inter,yan2013streamscan}. However, previous work fails to identify the deadlock issue in this global software barrier problem, thus no solution towards this issue. In contrast, {\simd} unveils, systematically analyzes, and resolves this problem. To avoid high register consumption, {\simd} further selectively fuse kernels via exploiting the special kernel launching patterns of graph algorithms. It is also important to mention existing work~\cite{wahib2014scalable} that only fuse kernels to barrier boundary. In comparison, {\simd} fuses kernels across barriers. \change{Our design can also benefit the popular Persistent Kernel~\cite{gupta2012study} designs which have been found suffer from deadlock issues when the occupancy exceed an unknown bound~\cite{maleki2016higher,hower2014heterogeneous}.} 

\section{Conclusion}
\label{sec-conclusion}

In this work, we propose {\simd}, a parallel graph computing framework that supports programming and processing of \textit{single instruction multiple, {complex}, data} on GPUs.
Specifically, the {Active-Compute-Combine} (ACC) model provides ease of programming to programmers, while  {just-in-time task management} and  {push-pull based kernel fusion} leverage the  opportunities for system-level optimization.  
Using {\simd}, a user can program a graph algorithm in tens of lines of code, while achieving significant speedup over the state-of-the-art.

\section*{Acknowledgment}
Hang Liu did part of this work as Graduate Research Assistant at the George Washington University. This work was partially supported by National Science Foundation CAREER award 1350766 and grants 1618706 and 1717774 at George Washington University. This research used resources from XSEDE and Amazon AWS research credits at University of Massachusetts Lowell.

\bibliographystyle{plainHang}
{
\bibliography{citation}
}


\end{document}